\documentclass[prd,twocolumn,showpacs,preprintnumbers,nofootinbib,prd,superscriptaddress,groupedaddress,10pt]{revtex4-1}

\usepackage{amsmath,amssymb}
\usepackage{amsfonts}
\usepackage{bm}
\usepackage{graphics}
\usepackage{float}
\usepackage{amsmath}
\usepackage{amssymb}
\usepackage[normalem]{ulem} 
\usepackage[mathscr]{euscript}
\usepackage{acronym}
\usepackage{float}
\usepackage[caption=false]{subfig}
\usepackage{lipsum}
\usepackage{mathtools}
\usepackage{multirow}
\usepackage{graphicx}
\usepackage[usenames,dvipsnames]{xcolor}
\usepackage[colorlinks, pdfborder={0 0 0}]{hyperref} 
\usepackage[nameinlink,capitalise]{cleveref}

\definecolor{LinkColor}{rgb}{0.75, 0, 0}
\definecolor{CiteColor}{rgb}{0, 0.5, 0.5}
\definecolor{UrlColor}{rgb}{0, 0, 0.75}
\hypersetup{linkcolor=LinkColor}
\hypersetup{citecolor=CiteColor}
\hypersetup{urlcolor=UrlColor}

\graphicspath{{fig}}

\usepackage{booktabs}

\DeclareMathAlphabet\mathpazo{OML}{zplm}{m}{it}
\def\thercsid{\relax}
\def\rcsid#1{\def\next##1#1{\def\thercsid{##1}}\next}
\rcsid$Id: QNM.tex,v  Doc No. $
\renewcommand{\today}{\number\day\space\ifcase\month\or
January\or February\or March\or April\or May\or June\or
July\or August\or September\or October\or November\or December\fi
\space\number\year}	

\usepackage{graphicx}
\usepackage[caption=false]{subfig}
\usepackage{xspace}

\usepackage{mathrsfs}

\newcommand{\beq}{\begin{equation}}
\newcommand{\eeq}{\end{equation}}
\newcommand{\bit}{\begin{itemize}}
\newcommand{\eit}{\end{itemize}}

\newcommand{\ddat}{\boldsymbol{\mathcal{D}}} 
\newcommand{\pp}{\mathcal{P}} 
\newcommand{\tth}{\boldsymbol{\vartheta}} 
\newcommand{\mdl}{\mathcal{H}}

\newcommand{\nn}{\boldsymbol{\mathcal{N}}}
\begin{document}
\title{Detectability of the subdominant mode in a binary black hole ringdown}


\author{Swetha Bhagwat}
\email{spbhagwa@syr.edu}
\affiliation{Department of Physics, Syracuse University, Syracuse NY 13244, USA}
\affiliation{Dipartimento di Fisica, “Sapienza” Universita di Roma, Sezione INFN Roma1, Piazzale Aldo Moro 5, 00185, Roma, Italy}

\author{Miriam Cabero}
\affiliation{Department of Physics, Princeton University, Princeton, NJ 08544, USA}

\author{Collin D. Capano} 
\affiliation{Max Planck Institute for Gravitational Physics (Albert Einstein Institute), Callinstrasse 38, D-30167 Hannover, Germany}
\affiliation{Leibniz Universit\"at Hannover, Welfengarten 1-A, D-30167 Hannover, Germany}

\author{Duncan A. Brown}
\affiliation{Department of Physics, Syracuse University, Syracuse NY 13244, USA}

\author{Badri Krishnan}
\affiliation{Max Planck Institute for Gravitational Physics (Albert Einstein Institute), Callinstrasse 38, D-30167 Hannover, Germany}
\affiliation{Leibniz Universit\"at Hannover, Welfengarten 1-A, D-30167 Hannover, Germany}

\begin{abstract}

 The ringdown is the late part of the post-merger signature emitted during the coalescence of two black holes and comprises of a superposition of quasi-normal-modes. Within general relativity, because of the no-hair theorems, the frequencies and damping times of these modes are entirely determined by the mass and angular momentum of the final Kerr black hole.  A detection of multiple ringdown modes would potentially allow us to test the no-hair theorem from observational data.  The parameters which determine whether sub-dominant ringdown modes can be detected are primarily the overall signal-to-noise ratio present in the ringdown signal, and on the amplitude of the subdominant mode with respect to the dominant mode. In this paper, we use Bayesian inference to determine the detectability of a subdominant mode in a set of simulated analytical ringdown signals. Focusing on the design sensitivity of the Advanced LIGO detectors, we systematically vary the signal-to-noise ratio of the ringdown signal, and the mode amplitude ratio in order to determine what kind of signals are promising for performing black hole spectroscopy. 


\end{abstract}


\maketitle 

\section{Introduction}
\label{sec:Intro}

The morphology of the gravitational wave signal from a binary black hole merger is well known.  Initially when the black holes are far apart, the signal is oscillatory with increasing amplitude and frequency; this part of the signal is well described by post-Newtonian theory.  As the black holes get closer and merge to form a single remnant black hole, the post-Newtonian description breaks down.  The amplitude reaches a maximum and then decreases as the remnant black hole approaches its equilibrium state, that of a Kerr black hole.  At some point after the merger, the remnant black hole spacetime is sufficiently close to its final equilibrium state that it can be well modelled as a linear perturbation of a Kerr black hole.  Power-law tails are expected at still later times, but these are likely too weak to be observable.  

The equation governing the perturbation of a Kerr black hole can be cast in the form of a radiative boundary-value problem similar to a Schr\"odinger equation (though with a non self-adjoint operator) with an effective potential depending on the mass $M$ and specific angular momentum $a$ of the black hole \cite{Teukolsky1,Teukolsky2,Teukolsky3,Vishveshwara:1970zz,Chandrasekhar:1985kt}.  Imposing boundary conditions which are purely outgoing at infinity and purely infalling into the black hole horizon leads to exponentially damped sinusoidal solutions, the quasi-normal-modes (QNMs).  For any given values of $M$ and $a$, the frequencies $f_{n\ell m}(M,a)$ and damping times $\tau_{n\ell m}(M,a)$ of the QNMs are determined by three quantum numbers $\ell,m,n$; $\ell$ and $m$ are the angular quantum numbers while $n$ denotes the overtones, i.e. number of zeroes of the radial part of the wave-function. 

If one were to observe a single QNM, then a knowledge of the mode indices $\ell,m,n$ would allow us to measure the mass and spin of the remnant black hole.  It is reasonable to assume that at sufficiently late times the least damped mode will dominate, but this may not be true closer to the merger.  Thus, it was found that the late time behavior of the first binary black hole merger detection, GW150914, is consistent with the $\ell=m=2, n=0$ quasi-normal-mode \cite{TheLIGOScientific:2016src}.
This question is of course closely connected to the issue of quantifying the time after which the remnant black hole can be treated perturbatively.  Studying the $n=0$ modes in numerical simulations of binary black hole mergers, it was suggested in \cite{Kamaretsos:2011um} that starting from a time $\sim 10GM/c^3$ after the merger, the gravitational wave signal is consistent with the quasi-normal-mode frequencies calculated from black hole perturbation theory (this is also consistent with the observational result in \cite{TheLIGOScientific:2016src}).  This is further suppported by an entirely different calculation, namely the decay of the horizon multipole moments \cite{Gupta:2018znn}; it is found that the decay rates of the horizon multipole moments become consistent with the quasi-normal-mode damping times roughly $10GM/c^3$ after the merger. See also \cite{MeMasha} for a quantitative study of how the near horizon geometry approaches a Kerr solution. 

Recent work indicates that higher overtones might contribute closer to the merger \cite{Giesler:2019uxc,2019arXiv191008708B}.  In fact, it turns out that the decay rates of the multipole moments calculated in \cite{Gupta:2018znn} are also consistent with the higher overtones closer to the merger. It is also possible to test general relativity by checking the consistency of $(M,a)$ between the ringdown and pre-merger portions of the signal \cite{Ghosh:2016qgn,2018PhRvD..97l4069C,TheLIGOScientific:2016src}.

The observation of \emph{multiple} ringdown modes would allow further tests of general relativity, and this is often referred to as black hole spectroscopy.  As first proposed in \cite{0264-9381-21-4-003}, verifying the consistency of $(M, a)$ measured from different modes allows us to test the no-hair theorem in general relativity.  We note that one could also use information from the full inspiral-merger-ringdown models as in \cite{Brito:2018rfr}.  While these are valid tests, more stringent tests, i.e. with fewer assumptions, are possible if no additional input based on earlier portions of the waveform are used.  This requires a detection of the sub-dominant ringdown mode purely from the post-merger portion of the signal. In this regard, the observation of higher overtones closer to the merger \cite{Isi:2019aib}, if fully confirmed, makes the prospects for black hole spectroscopy very promising.

From a data analysis perspective, the prospects of measuring a subdominant mode depend primarily on the overall signal-to-noise ratio (SNR) ($\rho_{RD}$) of all the ringdown modes, and the mode excitation amplitude $A_R$ of the subdominant mode. For instance, a nearly equal mass binary system like GW150914 does not excite the subdominant modes sufficiently and thus is not ideal for inferring multiple modes in the QNM spectrum. As a rule of thumb, the higher the asymmetry in the progenitor system, the lower is the $\rho_{RD}$ needed to detect subdominant modes. This question, (along with the related issue of resolving nearby frequencies and damping times) was previously studied using Fisher matrix approximations (see e.g. \cite{bertiparam,Fh1}).  More recently, especially in the era of gravitational wave detections, Bayesian parameter estimation techniques have proven to be very effective, and several toolkits exist specifically tailored towards gravitational wave astronomy (see e.g. \cite{Veitch:2014wba,Ashton:2018jfp,Biwer:2018osg}).  In this paper we study the detectability of the sub-dominant ringdown mode using these Bayesian inference techniques for the case when there are two potentially detectable modes. Specifically, we study the effect of varying $A_{R}$ and $\rho_{RD}$ on the recovery of the ringdown parameters. We assume that the underlying theory of gravity is standard general relativity, and compute the frequencies and damping times as dictated by linear perturbation theory for a given Kerr black hole \cite{1985RSPSA.402..285L,2006PhRvD..74j4020B}.

This paper is organized as follows: in section \ref{sec:theory} we describe the ringdown waveform and the details of injections.  In section \ref{sec:PE_theory}, we describe the setup for our Bayesian inference procedure. In section \ref{sec:result} we present our results and finally discuss its implications in \ref{sec:conclusion}.

\section{The ringdown waveform}
\label{sec:theory}

To describe the GW ringdown signal, we begin with a general plane gravitational wave.  Let $h_{\mu\nu}$ be a symmetric transverse-traceless metric perturbation tensor and construct a orthonormal wave frame with basis vectors $(\hat{X},\hat{Y},\hat{Z})$ with $\hat{Z}$ being the wave propagation direction.  The metric perturbation has two independent degrees of freedom and can be written as 
\begin{equation}
    h_{\mu\nu} = h_+(\mathbf{e}_+)_{\mu\nu} + h_\times (\mathbf{e}_\times)_{\mu\nu}
\end{equation}
where 
\begin{equation}
    \mathbf{e}_+ = \hat{X}\otimes\hat{X} - \hat{Y}\otimes\hat{Y}\,,\quad \mathbf{e}_\times = \hat{X}\otimes\hat{Y} + \hat{Y}\otimes\hat{X}\,.
\end{equation}
We are allowed to make further rotations in the $\hat{X}-\hat{Y}$ plane, and we can choose a preferred frame such that $h_{+,\times}$ as functions of time $t$ have the form
\begin{equation}
    h_+(t) = A_+(t)\cos\Phi(t) \,,\quad h_\times(t) = A_\times(t)\sin\Phi(t)\,.
\end{equation}
where $A_{+,\times}$ are slowly varying amplitudes and $\Phi(t)$ is a rapidly varying phase. 

The response of a detector depends linearly on $h_{+,\times}$ and also on the three Euler angles which relate the wave-frame $(\hat{X},\hat{Y},\hat{Z})$ to the detector frame.  These three angles are related to the sky-position of the source in the detector frame represented by the right ascension $\alpha$ and declination $\delta$, and on the orientation of the preferred $(\hat{X},\hat{Y})$ frame represented by the so-called polarization angle $\psi$.  In the limit when the wavelength of the signal is much larger than the size of the detector, it can be shown that the strain $h(t)$ observed by a detector is 
\begin{equation}
    h(t) = F_+(\alpha,\delta,\psi)h_+(t) + F_\times(\alpha,\delta,\psi)h_\times(t)\,.
\end{equation}
For a binary system, or for a perturbed Kerr black hole, there is a natural axis of rotation.  In the case of a binary system this axis is the normal to the orbital plane, and in the case of a perturbed Kerr black hole, this is the spin direction of the Kerr black hole.
There are three additional angles relating the waveframe to the source frame.  These are related to the direction to the detector in the source frame via the inclination angle $\iota$ (the angle between the source axis and the line of sight to the source in the detector frame) and a reference orbital phase $\varphi$.  We could, if desired, introduce an additional polarization angle of the orientation of the $(\hat{X},\hat{Y})$ in the source frame, but it is conventional to absorb this in the angle $\psi$ defined above.  

Specializing now to ringdown waveforms, $h_{+,\times}$ can be expanded as a superposition of damped sinusoids
\begin{equation}
    h_{+}+ih_\times = \sum_{\ell,m,n} {}_2Y_{\ell m}(\iota,\varphi) A_{\ell m n}e^{i(\omega_{\ell m n}t + \phi_{\ell m n})}\,.
\end{equation}
Here ${}_2Y_{\ell m}$ are the spin-weighted spherical harmonics \cite{Goldberg:1966uu,GMS}, $\omega_{\ell m n}$ is the complex ringdown frequency, $\phi_{\ell m n}$ the initial phase, and $A_{\ell m n}$ the (real) mode amplitude.  For a perturbed Kerr black hole, it is actually natural to use the spin-weighted spheroidal harmonics \cite{Chandrasekhar:1985kt} instead of the spin-weighted spherical harmonics ${}_2Y_{\ell m}$, but this is a good approximation up to moderately high spins (see e.g. \cite{Berti:2014fga}).  We use the method of continued fractions developed in \cite{1985RSPSA.402..285L} to calculate the frequencies and damping times.  

In this paper we shall restrict ourselves to two modes, and the one with the lower signal-to-noise ratio will be called subdominant.  Furthermore, for simplicity, we shall consider only the $\ell = m = 2, n=0$ and $\ell=m=3, n=0$ modes.  Given that both modes considered have $n=0$ we shall drop the overtone index in the frequencies and damping times.  These frequencies and damping times are of course determined by the mass $M_f$ and specific angular momentum $a_f$ of the final black hole.  The $\ell=m=2$ mode is taken to be the dominant one with amplitude $A_{22}$. The amplitude $A_{33}$ will be parameterized via the amplitude ratio $A_R := A_{33}/A_{22}$. Thus, the waveform model we consider in this paper is fully described by the parameters $\{M_{f}, a_{f}, A_{22}, A_{R}, \phi_{22}, \phi_{33} \} $


\section{On the Parameter Estimation and its implementation}
\label{sec:PE_theory}

Consider a network of $N$ detectors labeled by indices $i,j=1\ldots N$.  Let the data from the $i^{\rm th}$ detector be $d_i(t)$ and denote the collection of $N$ time-series $\{d_i(t)\}$ by $\ddat$.  A binary system is parameterized by intrinsic parameters such as masses, spins, etc. which affect the phase evolution of the signal, and also by extrinsic parameters such as sky position, luminosity distance, coalescence time, etc. which affect the slowly varying amplitude.  Let $\tth$ denote all of these parameters collectively.  The signal model is then written as $h(t;\tth)$, and by considering the responses of the detectors in the network we obtain the signal $h_i(t;\tth)$ as seen by the $i^{\rm th}$ detector, denoted collectively as $\mdl$.  Given the presence of a signal, the goal of any parameter estimation procedure is to determine the most likely values of $\tth$ with suitable error-bars.  

The framework of Bayesian inference provides a general framework for determining probability distributions for the parameters $\tth$ known as the posterior distributions $p(\tth|\ddat,\mdl)$.  Given a $\mdl$, one has expectations on the distribution of parameter values before performing an observation \cite{BayesBooks1,BayesBooks2} encoded in a probability density function called the \textit{prior},  $\pp(\tth|\mdl)$. Once the observation is performed and the data set is obtained, one updates the priors with information obtained from this observation. This input is encoded in the \textit{Likelihood function} $\pp(\ddat|\tth,\mdl)$.   The posterior probability density function $\pp(\tth|\ddat,\mdl)$ for the parameters $\tth$ is given by \cite{BayesBooks1,BayesBooks2}, 
\begin{align}
\label{eq:bayes_thrm}
\pp(\tth|\ddat,\mdl) = \frac{\pp(\tth|\mdl) \pp(\ddat|\tth,\mdl)}{\pp(\ddat|\mdl)}.
\end{align}
Here, $\pp(\ddat|\mdl)$ is the \textit{evidence} and serves as a normalization factor. 

The likelihood function $\pp(\ddat|\tth,\mdl)$ depends on both the signal and the nature of noise present in the data. Let the time-series data from the detector contain a GW signal $\mdl$ embedded in the detector noise $\nn$, i.e $\ddat= \mdl+\nn$. If the noise model is Gaussian and stationary, the likelihood function, $\pp(\ddat|\tth,\mdl)$, can be written as 
\begin{align}
\label{eq:likelihood}
\pp(\ddat|\tth,\mdl) \propto e^{-\frac{1}{2} \left\langle \nn|\nn  \right\rangle}= e^{-\frac{1}{2} \left\langle \ddat - \mdl|\ddat - \mdl  \right\rangle}.
\end{align}
Here, $\left\langle .|. \right\rangle$ denotes an inner product in the space of signals.  In the case when there are no correlations between data from different detectors, the inner product is of the form
\begin{equation}
    \left\langle \mathbf{X}|\mathbf{Y} \right\rangle = \sum_{i=1}^N\left\langle x_i|y_i \right\rangle_i\,.
\end{equation}
For the $i^{\rm th}$ detector, the inner product is easiest to express in the frequency domain.  Let $\widetilde{x}_i(f)$ and $\widetilde{y}_i(f)$ be the Fourier transforms of $x_i(t)$ and $y_i(t)$ respectively, and let the single sided power spectral density of the noise be $S_n^{(i)}(f)$.  The inner product is then
\begin{equation}
    \left\langle x_i|y_i \right\rangle_i = 2\textrm{Re}\int_0^\infty \frac{\widetilde{x}_i^\star(f)\widetilde{y}_i(f)}{S^{(i)}_n(f)}\,df\,.
\end{equation}
The form of the prior distributions, $\pp(\tth|\mdl)$, is a choice that one has to make and there is no unique way to pick it. With the intention of extracting maximum information from the data itself, a non-informative prior is used in this study. 

All the information about the distribution of the estimated parameters is contained in the landscape of $\pp(\tth|\ddat,\mdl)$ and therefore, the goal of a scheme using Bayesian parameter estimation is to sample the parameters space of $\tth$ and construct the distribution $\pp(\tth|\ddat,\mdl)$  \footnote{In most cases where one is just interested in estimating the parameter values for $\tth$, $\pp(\tth|\ddat,\mdl)$ is calculated up to a normalization factor. One need not compute the evidence to estimate the parameters of the model. Calculating the evidence is computationally challenging, especially when the parameter space spanned by $\tth$ is large.}. In practice, these posterior distributions are computed by sampling \cite{sampling} the allowed parameter space using routines for sampling employ an algorithm called the Markov Chain Monte Carlo (MCMC)  \cite{MCMC1,MCMC2,MCMC3}. In our study, we employ a parallel tempered sampling algorithm called \texttt{emcee\_pt} which uses multiple temperature 'walkers' performing random walks to explore the parameter space without getting stuck in local minima. 

We perform a full Bayesian parameter estimation using the \texttt{PyCBC} package \cite{Biwer:2018osg} to produce the posterior distribution for the 6 ringdown parameters listed above. We use the inbuilt implementation of the \texttt{emcee\_pt} ensemble sampler to perform the parallel tempered Markov chain Monte Carlo (MCMC) operation. The technical details of this algorithm are presented in \cite{2013PASP..125..306F}. We use $38$ inverse-temperatures chains to sample the parameter space. For each temperature chain, we use 200 walkers to explore the space. We use an analytical model of the advanced LIGO sensitivity curve, named \texttt{Zero-detuned-high power} (ZDHP) noise curve\footnote{\texttt{https://dcc.ligo.org/LIGO-T1800044/public}}, for calculating the likelihood function at each sampled point.  

We perform the parameter estimation for the 6 ringdown parameters listed earlier: $\{M_{f}, a_{f}, A_{22}, A_{R}, \phi_{22}, \phi_{33} \} $.  All of the other extrinsic parameters are kept fixed.    In a more realistic case, these other extrinsic parameters would have to be provided independently.  This could be either from an unmodeled search using minimal assumptions (see e.g. \cite{Abbott:2019prv,Abbott:2016ezn}), or from a full inspiral-merger-ringdown modelled search.  In either case, the assumption is that these extrinsic parameters by themselves only depend on the propagation of gravitational radiation in our universe, and not on intrinsic properties of the source, such as the validity of the no-hair theorem.  

For simplicity, we perform these injections in zero noise. Note that zero noise is a realisation of Gaussian noise and therefore any assumptions during the PE that rely on the nature of noise being Gaussian still remain valid. A more detailed work of similar nature needs to be performed in the presence of detector noise to understand how it influences the analysis in a realistic case, but this is beyond the scope of our current study. 

We use the non-informative priors summarised below:
\begin{itemize}
\item $M_f$: Uniform between $[50,100]M_\odot$.
\item $a_f$: Uniform between $[-0.99,0.99]$.
\item $A_{22}$: Log-Uniform between $[10^{-25},5\times 10^{-20}]$.
\item $A_R$: Uniform between $[0,0.5]$.
\item $\phi_{22}$ and $\phi_{33}$: Uniform between $[0,2\pi]$.
\end{itemize}
A Log-uniform prior on $A_{22}$ is appropriate since it is an amplitude thereby setting the scale of the ringdown signal.  This choice also ensures a better sampling of the smaller amplitudes. On the other hand, since $A_R$ is a ratio of amplitudes, a uniform prior is appropriate.

\section{Results}
\label{sec:result}

\subsection{Parameters of the injected signals}

We simulate analytical ringdown signals and inject them into zero noise fake data. As described earlier, the injections used in this study contain only two QNMs, namely $\ell=m=2,3, n=0$, with the frequencies and damping times consistent with those for the Kerr quasi-normal modes for some value of the mass and spin.  Unlike a real BBH signal, there is no non-linear merger physics in these injections.  The injections correspond to a black hole with $\{ M_{f} = 70 M_\odot,  a_{f} = 0.65 \}$, somewhat similar to those for GW150914.  We consider only the two Advanced LIGO detectors at their design sensitivity.

The values of the extrinsic parameters for all of our injections are:  
\begin{itemize}
    \item Inclination angle: $\iota = 0.7\,$rad.
    \item Right ascension and declination: $\alpha = 2.2\,\textrm{rad}, \delta= -1.24\,\textrm{rad}$.
    \item Polarization angle: $\psi = 0.3\,$rad .
    \item Initial phases: $\phi_{22} = 0, \phi_{33} = 1\,$rad.
\end{itemize}
These are arbitrary choices and we do not expect these to affect our results generically.  More importantly, we consider 16 combinations of the optimal ringdown SNR $\rho$ and the mode amplitude ratio $A_R = A_{33}/A_{22}$:
\begin{itemize}
    \item $\rho_{RD} = \{ 15, 20, 25, 30 \}$.
    \item $A_R = \{0, 0.1, 0.2, 0.3 \} $.  
\end{itemize}
For moderate mass ratios, we note that the $\ell=m=3$ mode has previously been found to be the strongest sub-dominant mode
(see e.g. \cite{MeSpec,2014PhRvD..90l4032L,2019arXiv190108516B}).  Furthermore, again for moderate values of the mass ratio, these studies show that it is not unreasonable to have $A_R = \mathcal{O}(10^{-1})$ which justifies the values for $A_R$ listed above.

\subsection{Detectability of the sub-dominant mode}




Figure 3 of \cite{bertiparam} presents the effect of the choice of $\iota$ on the observed amplitudes of QNM; we note that our choice of $\iota = 0.7\,$ rad is fairly favourable for viewing the subdominant mode. 

 \begin{figure}
 \subfloat{\includegraphics[width=0.45\textwidth]{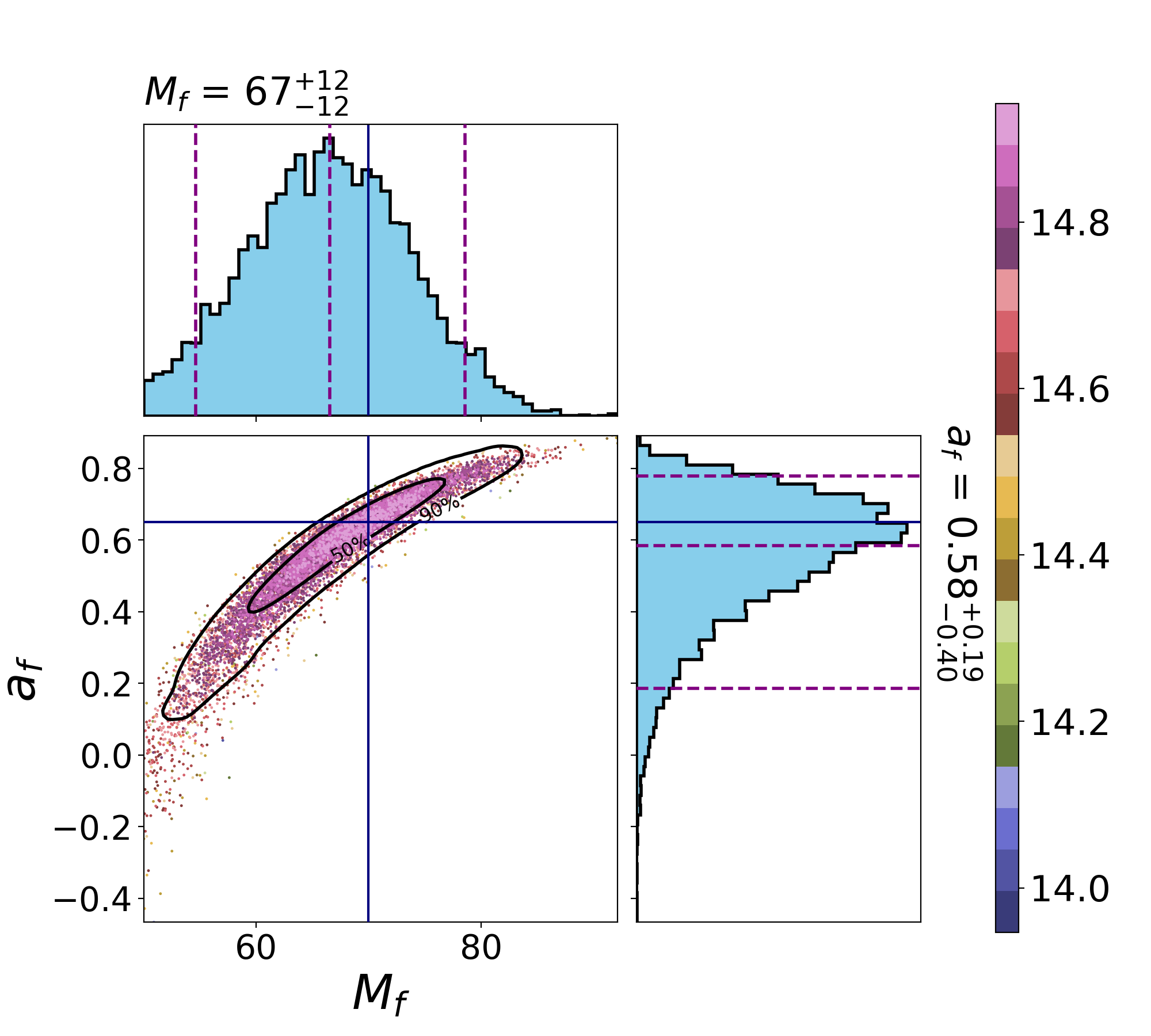} }
\caption{Recovery of final mass $M_{f}$ and final spin $a_{f}$ for SNR = 15 and $A_{R} =0.1$. In all the cases reported in this study we confirm that the $90 \%$ credible interval in the posterior distribution of the mass and spin of the final BH contains the injected parameter values. As an illustration we present the posterior distribution for final mass $M_{f}$ and spin $a_{f}$ corresponding to the injection with $A_{R}=0.1$ and $\rho_{RD}=15$.}
\label{fig:Mf-and-af}
\end{figure}

\begin{figure}
\subfloat{\includegraphics[width=0.25\textwidth]{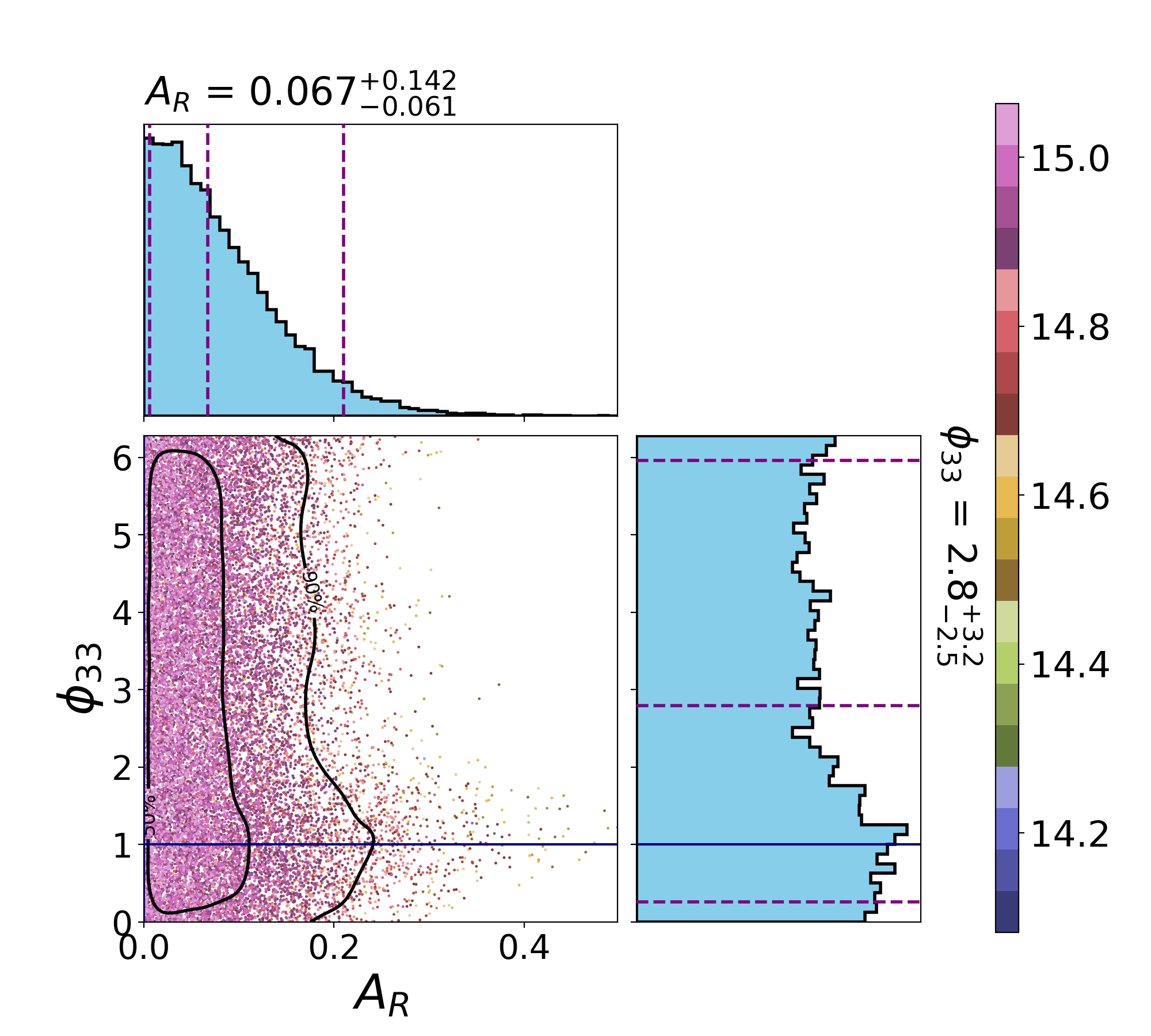} }
\subfloat{\includegraphics[width=0.25\textwidth]{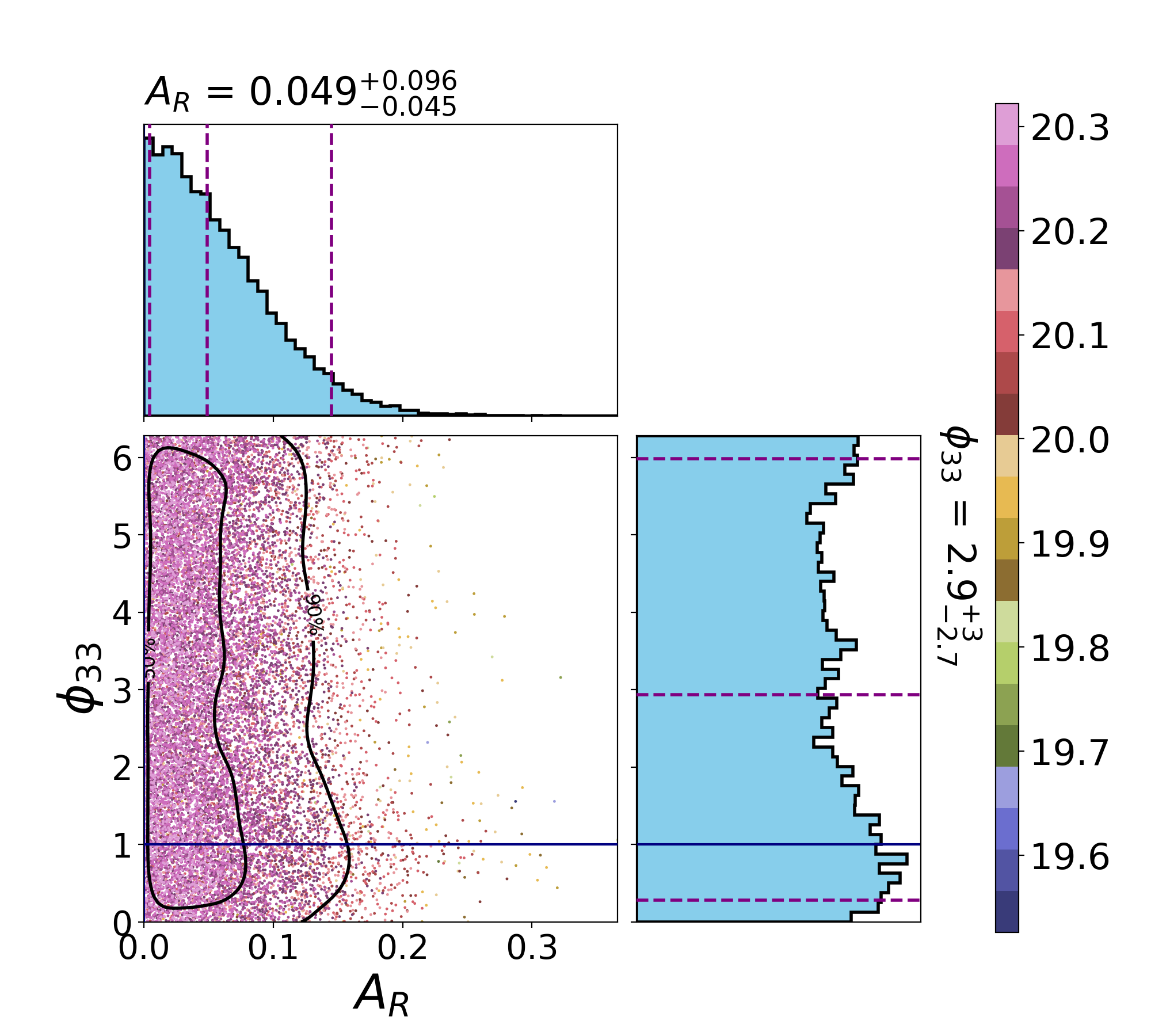} }  \\[-0.3cm]
\subfloat{\includegraphics[width=0.25\textwidth]{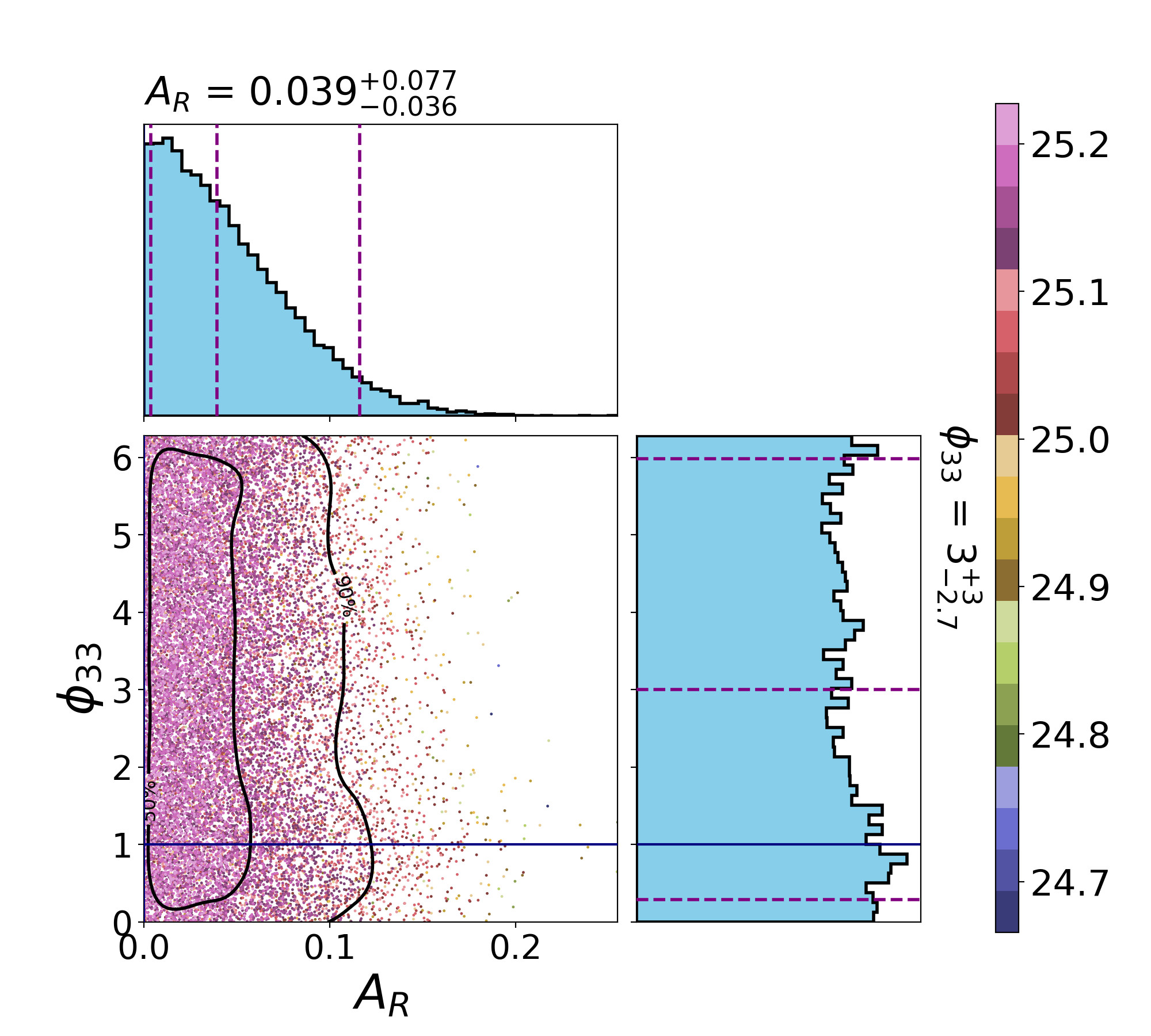} }
\subfloat{\includegraphics[width=0.25\textwidth]{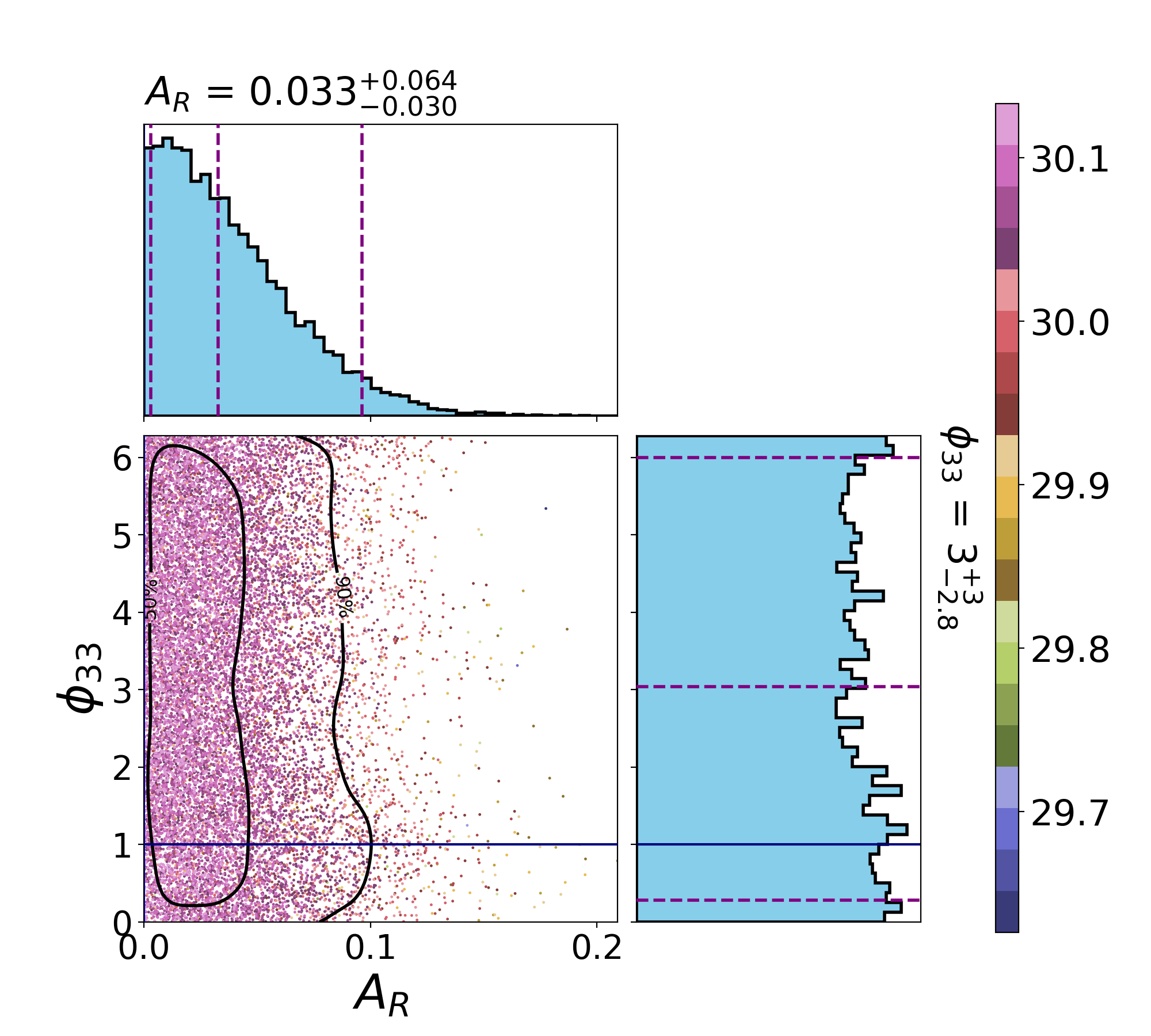} }
\caption{Posterior distributions for the amplitude ratio and phase of the subdominant mode in the null result test. The red lines indicate the injected values. The injected ringdown only contains the dominant mode, but is recovered by a template family that has two modes. The optimal SNR of the injection is (top left) $\rho_{RD} = 15$, (top right) $\rho_{RD} = 20$, (bottom left) $\rho_{RD} = 25$, and (bottom right) $\rho_{RD} = 30$.}
\label{fig:amp_ratio_0.0}
\end{figure}

\begin{figure}
\subfloat{\includegraphics[width=0.25\textwidth]{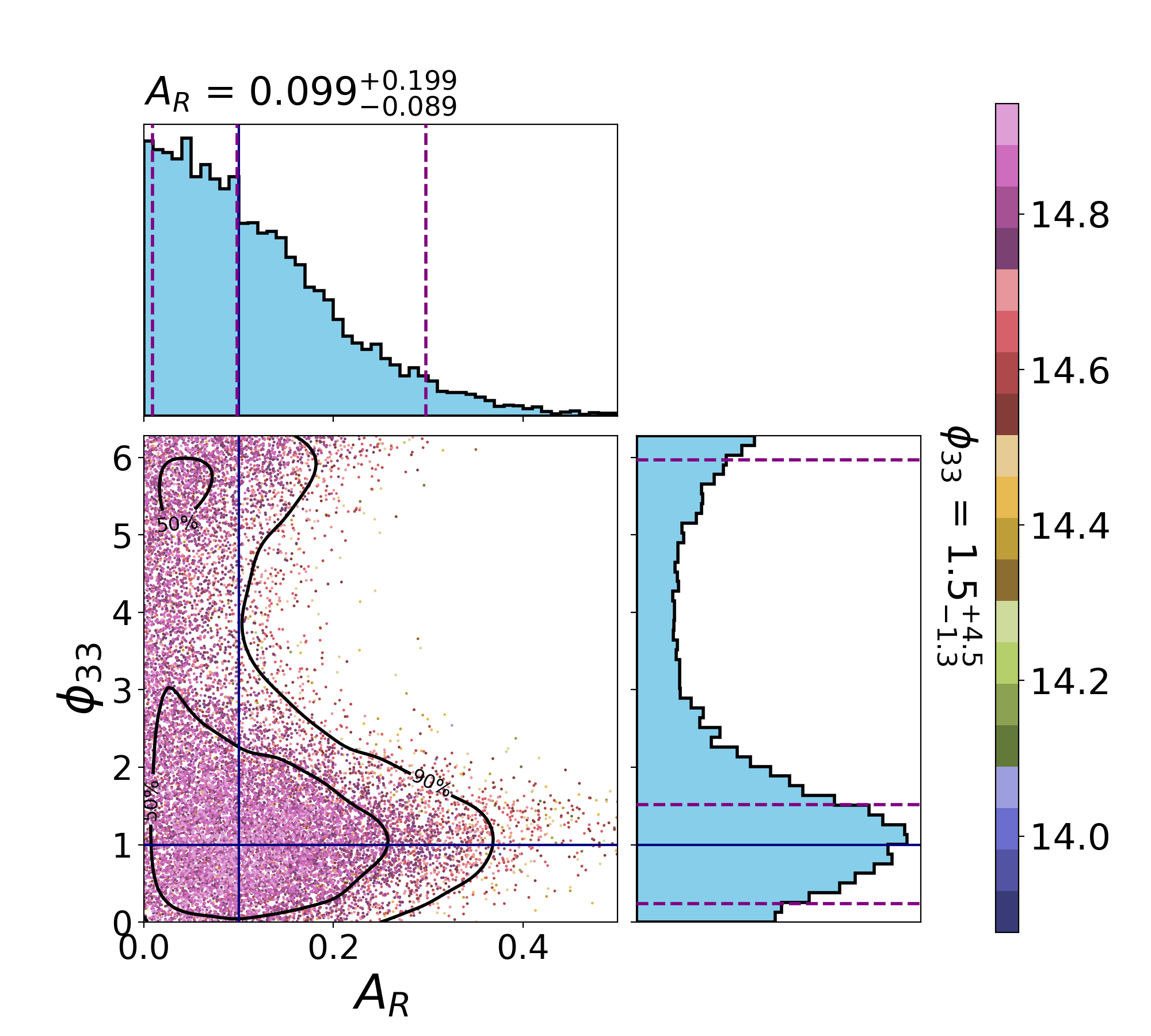} }
\subfloat{\includegraphics[width=0.25\textwidth]{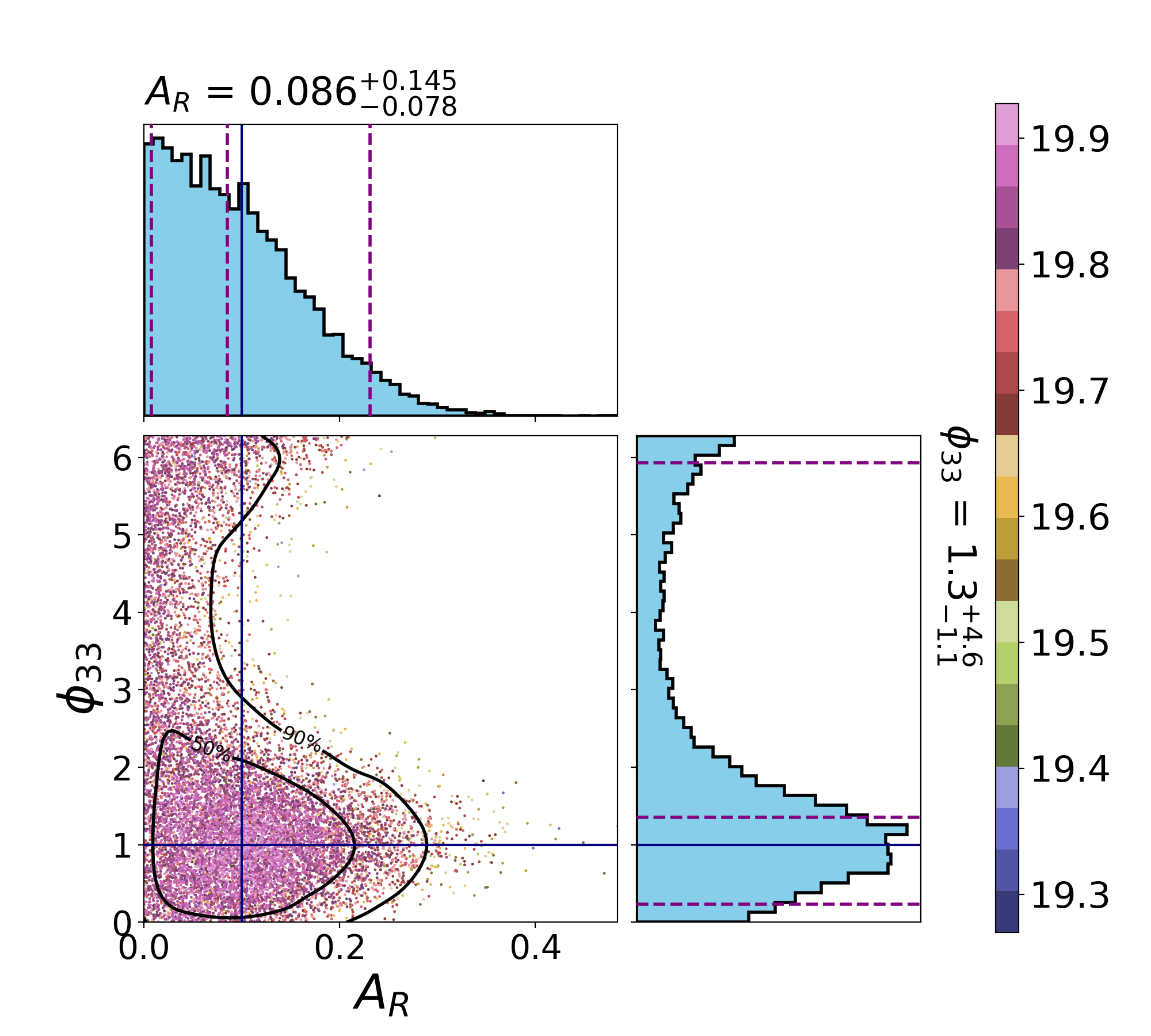} } \\[-0.3cm]
\subfloat{\includegraphics[width=0.25\textwidth]{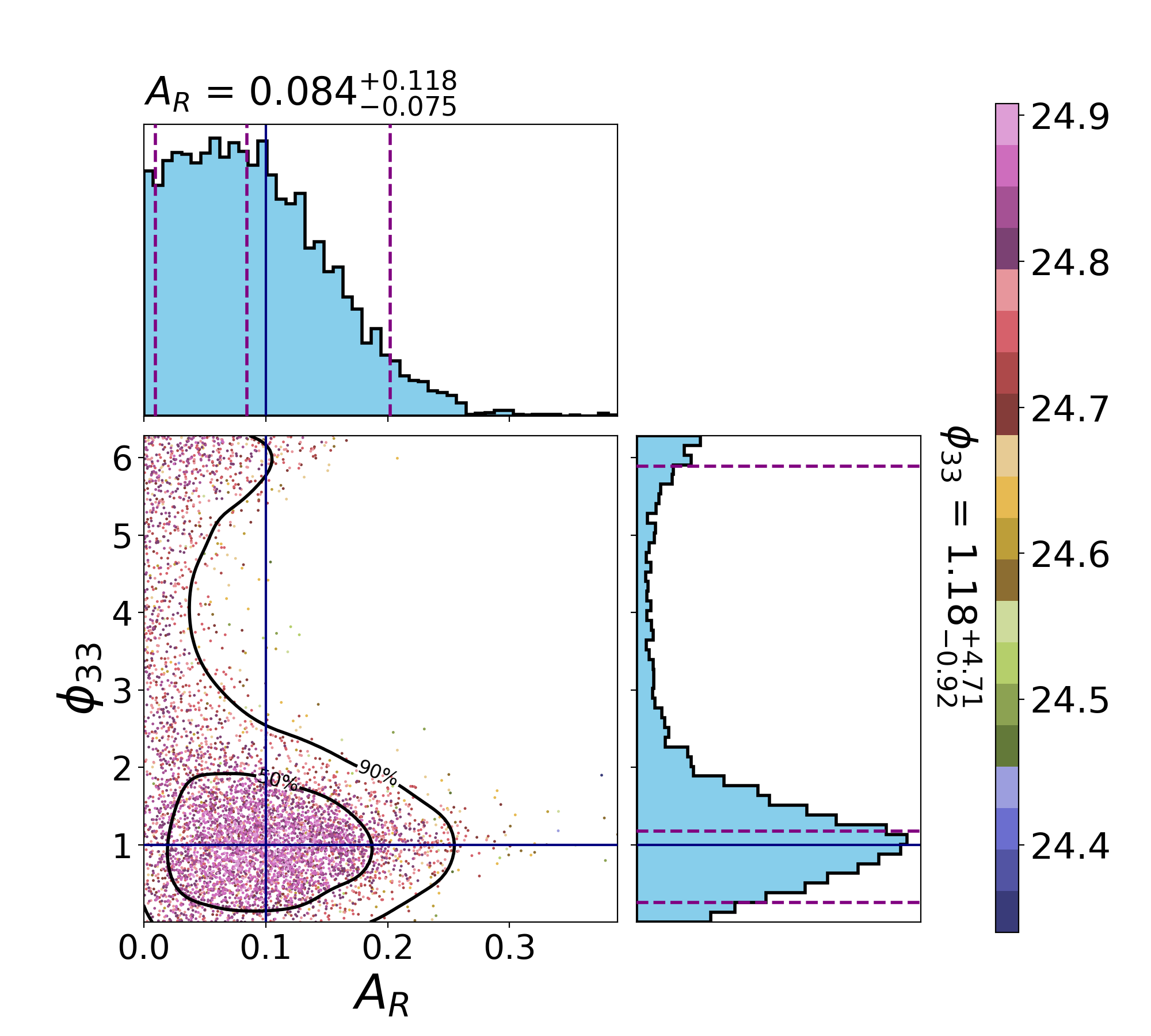} }
\subfloat{\includegraphics[width=0.25\textwidth]{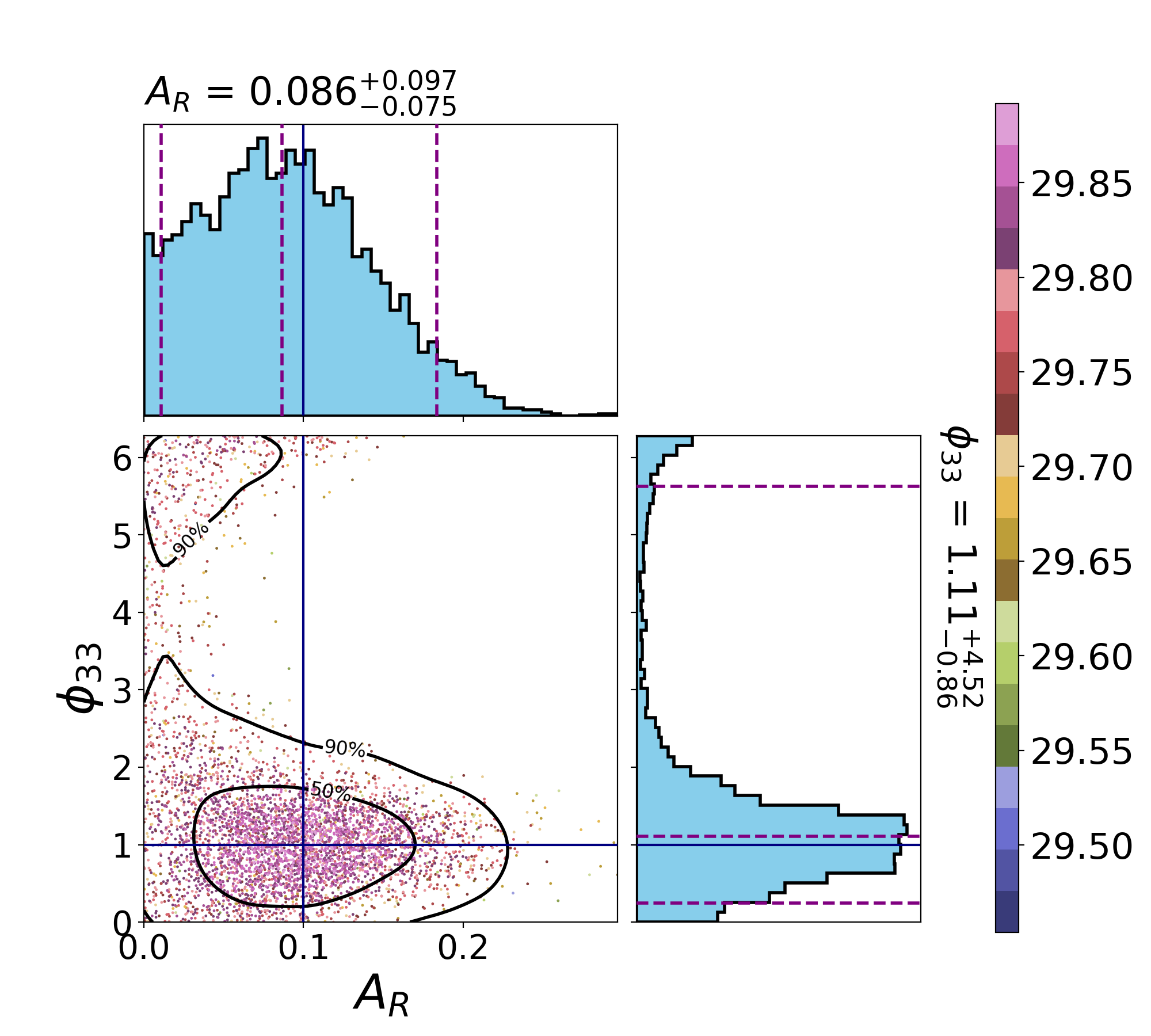} }
\caption{Posterior distributions for the amplitude ratio and phase of the subdominant mode in the $A_R = 0.1$ case. The red lines indicate the injected values. The optimal SNR of the injection is (top left) $\rho_{RD} = 15$, (top right) $\rho_{RD} = 20$, (bottom left) $\rho_{RD} = 25$, and (bottom right) $\rho_{RD} = 30$.}
\label{fig:amp_ratio_0.1}
\end{figure}

\begin{figure}
\subfloat{\includegraphics[width=0.25\textwidth]{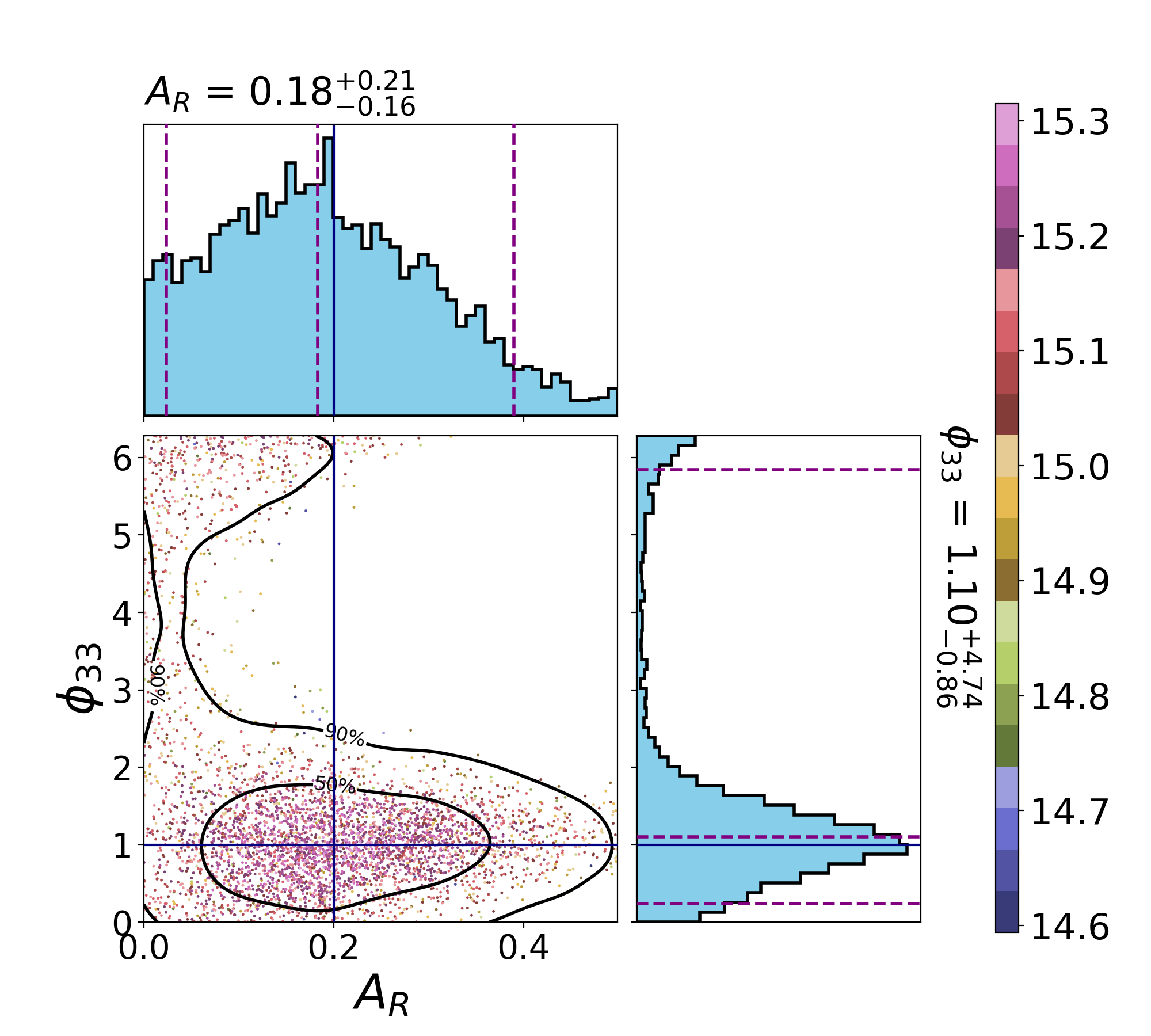} }
\subfloat{\includegraphics[width=0.25\textwidth]{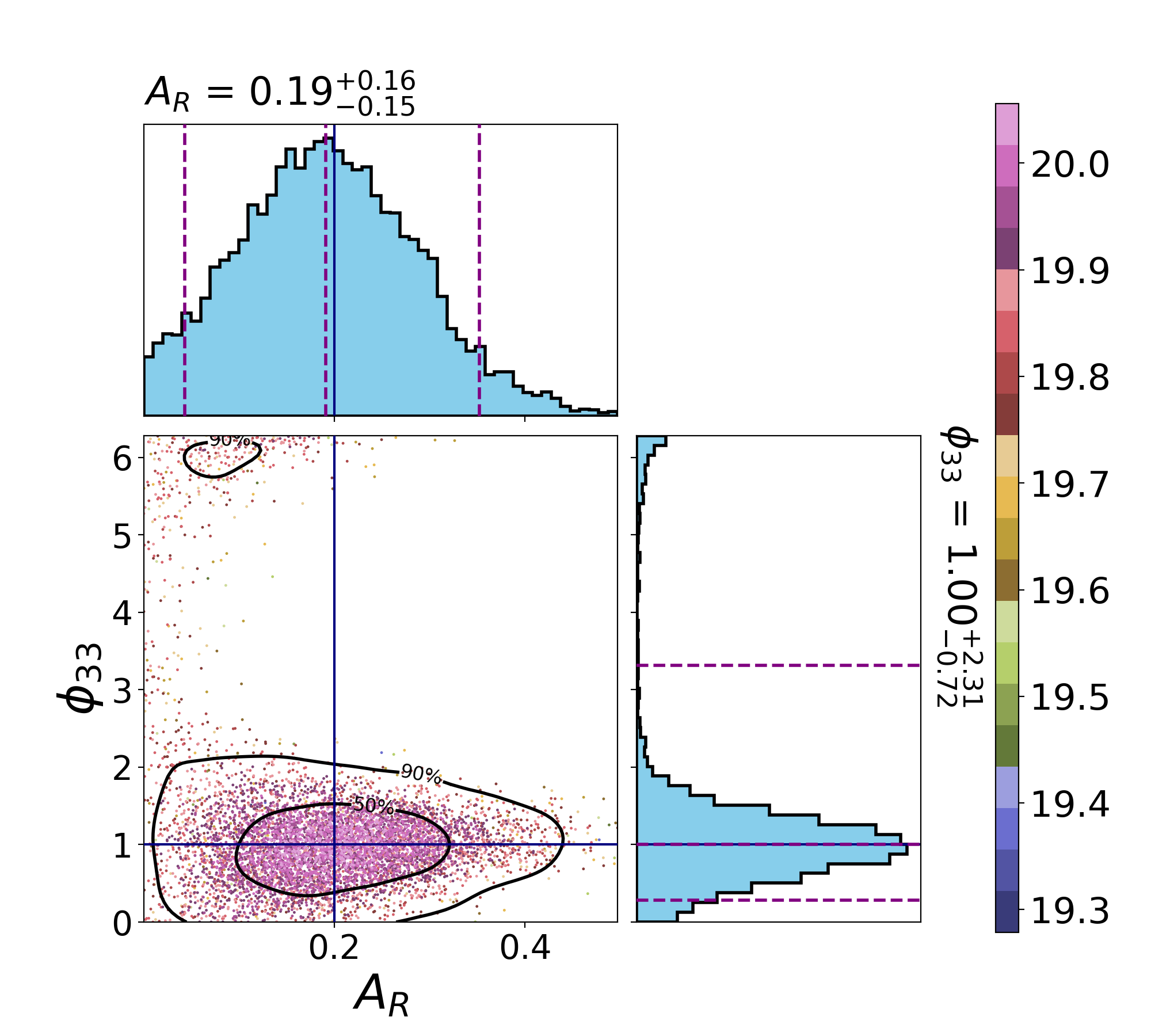} }  \\[-0.3cm]
\subfloat{\includegraphics[width=0.25\textwidth]{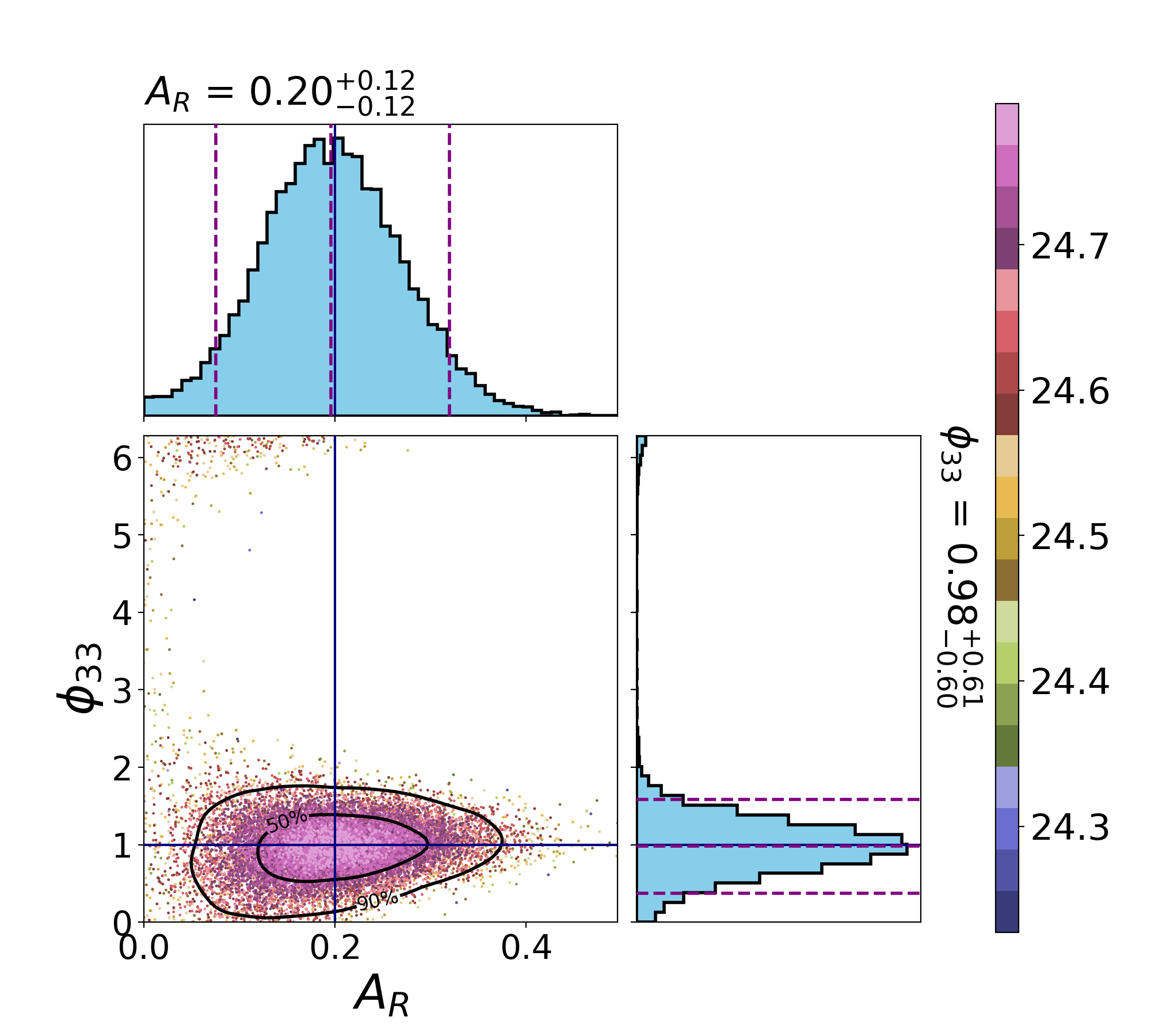} }
\subfloat{\includegraphics[width=0.25\textwidth]{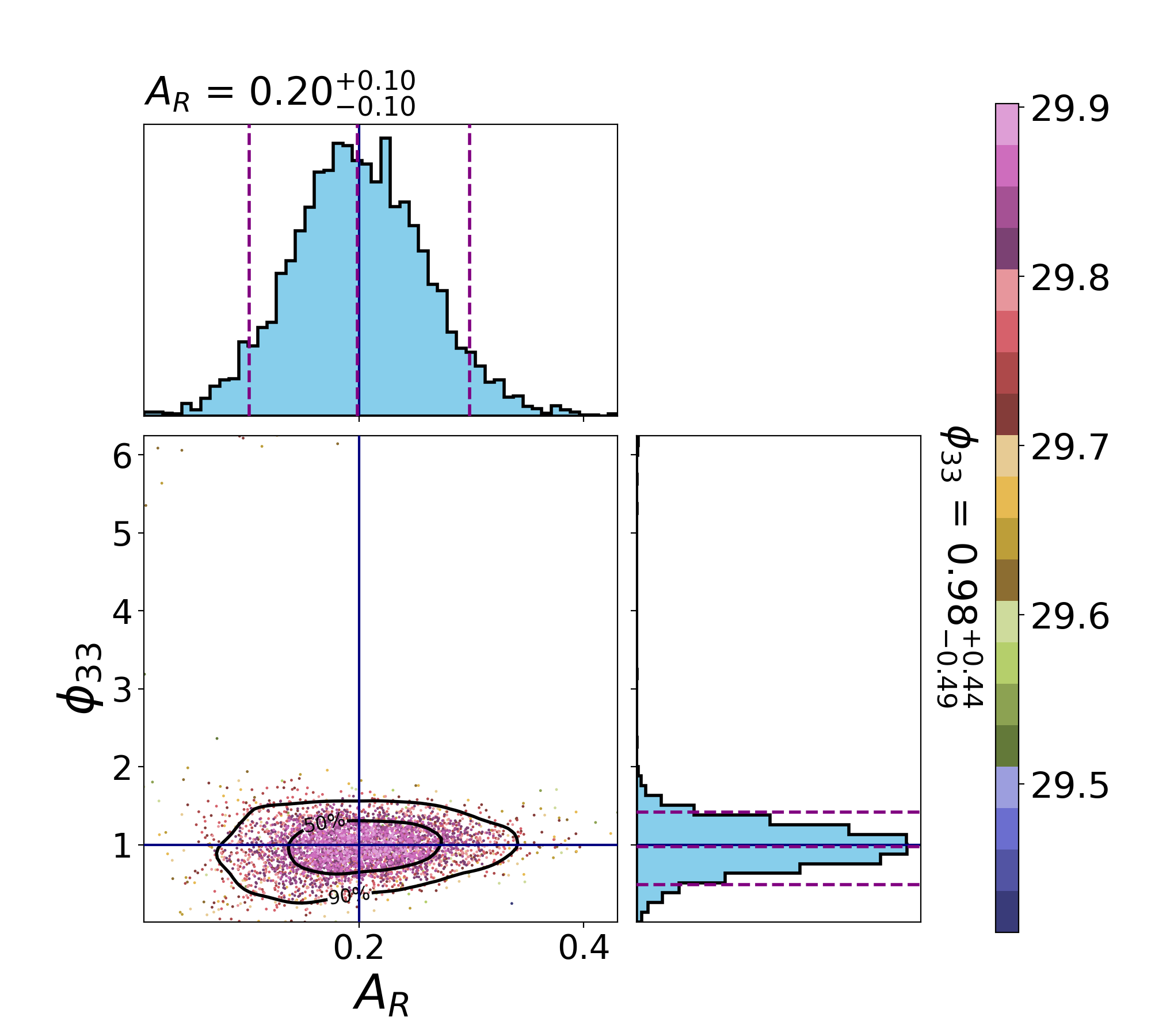} }
\caption{Posterior distributions for the amplitude ratio and phase of the subdominant mode in the $A_R = 0.2$ case. The red lines indicate the injected values. The optimal SNR of the injection is (top left) $\rho_{RD} = 15$, (top right) $\rho_{RD} = 20$, (bottom left) $\rho_{RD} = 25$, and (bottom right) $\rho_{RD} = 30$.}
\label{fig:amp_ratio_0.2}
\end{figure}
 
 \begin{figure}
\subfloat{\includegraphics[width=0.25\textwidth]{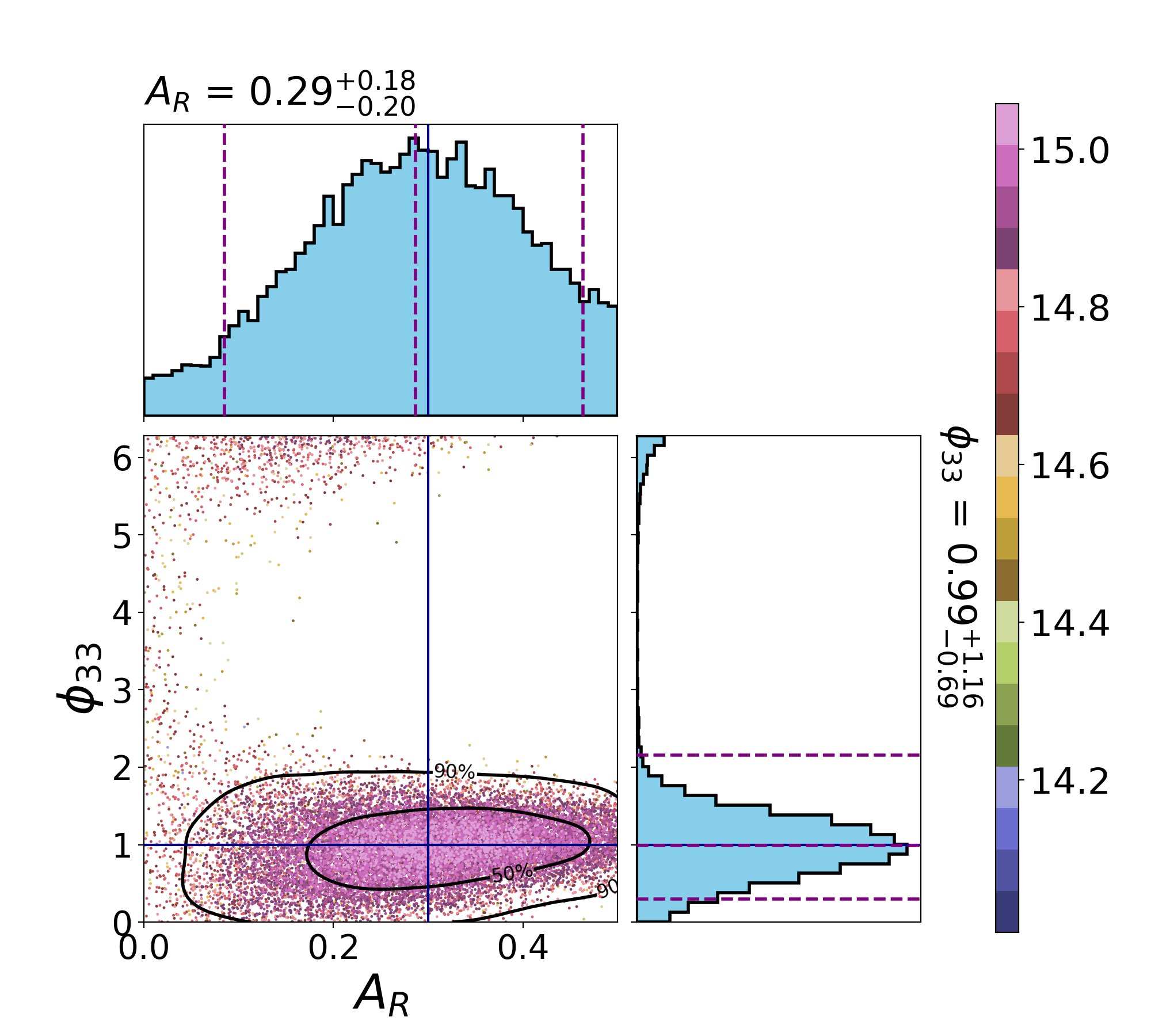} }
\subfloat{\includegraphics[width=0.25\textwidth]{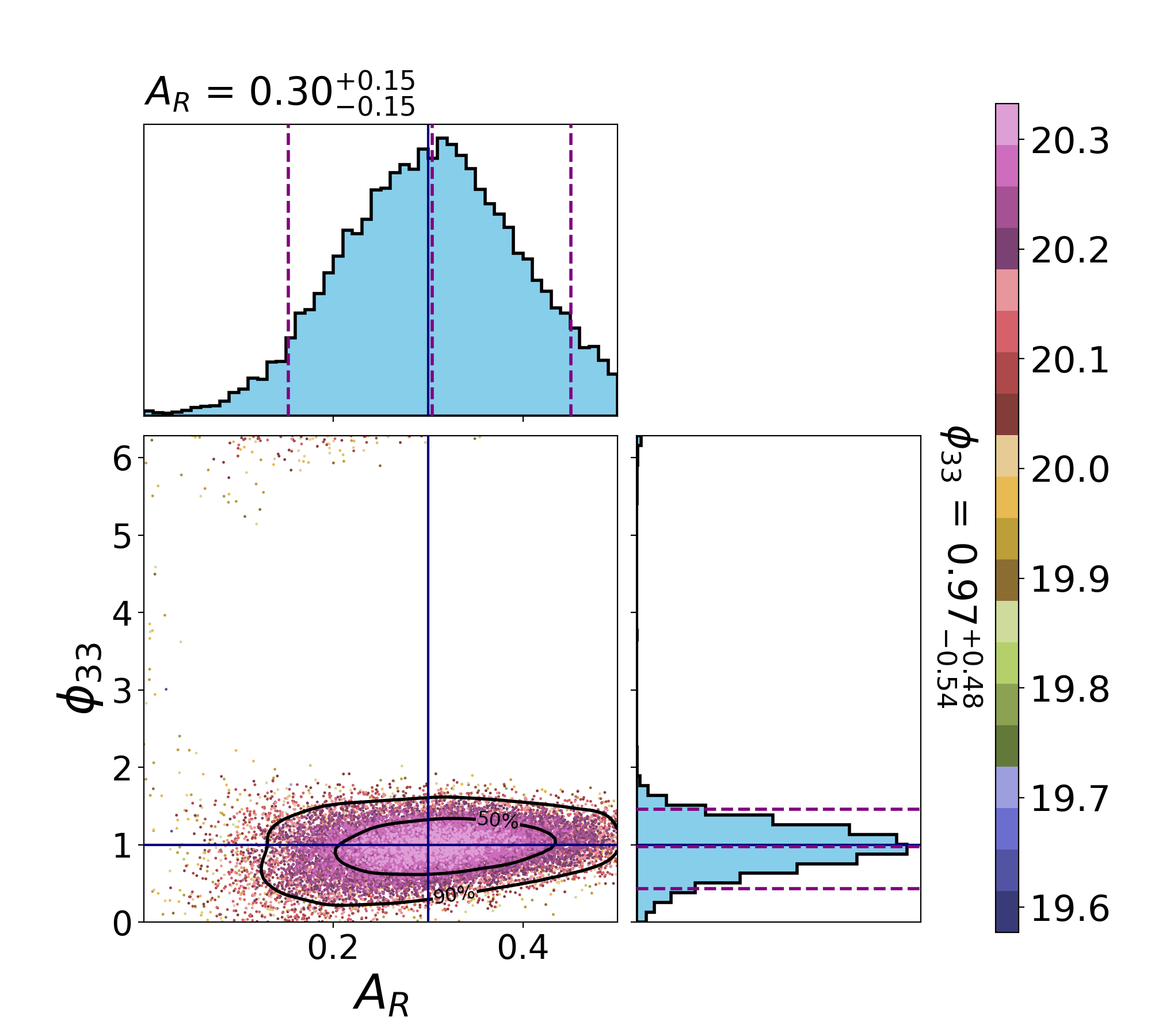} }  \\[-0.3cm]
\subfloat{\includegraphics[width=0.25\textwidth]{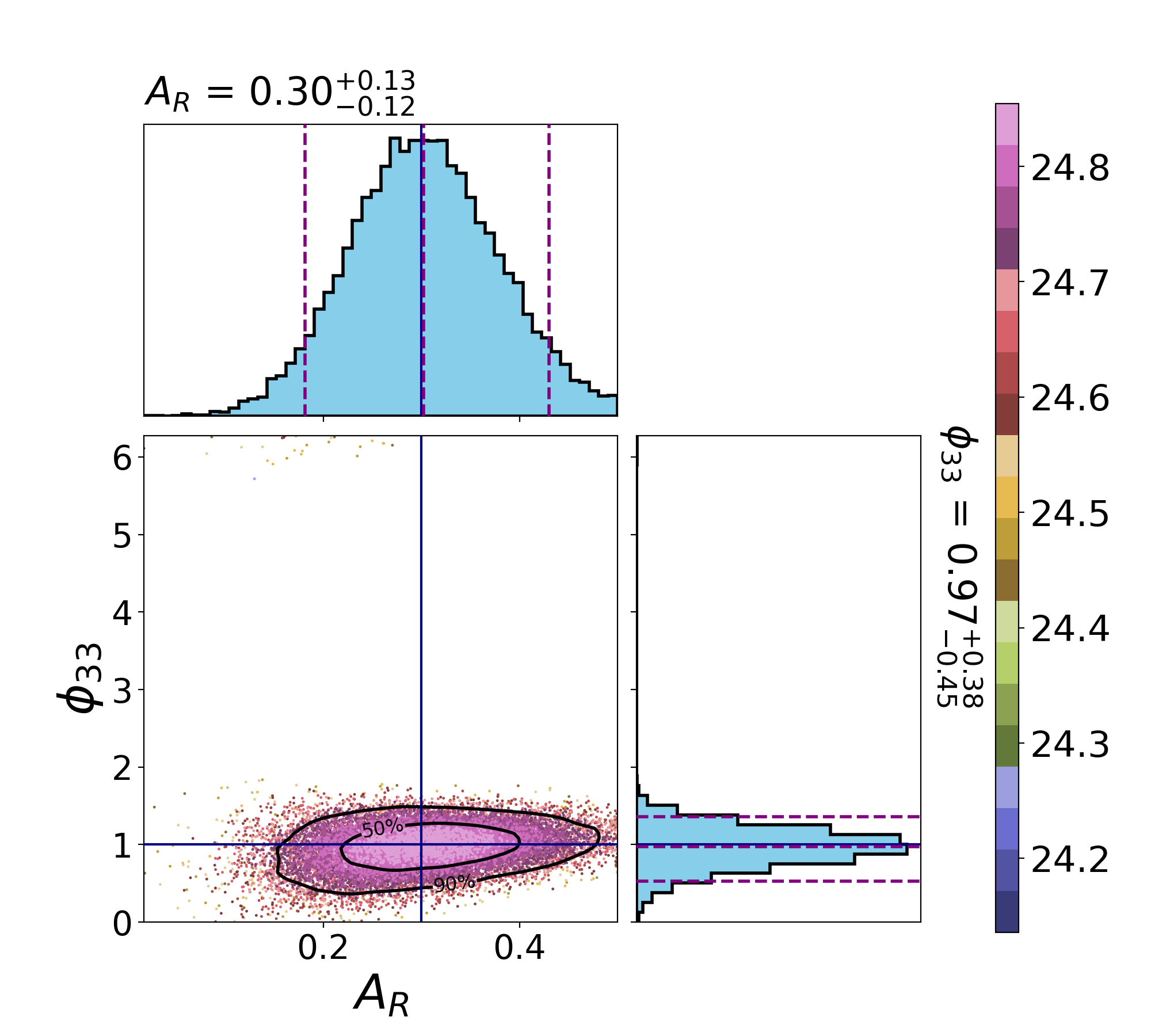} }
\subfloat{\includegraphics[width=0.25\textwidth]{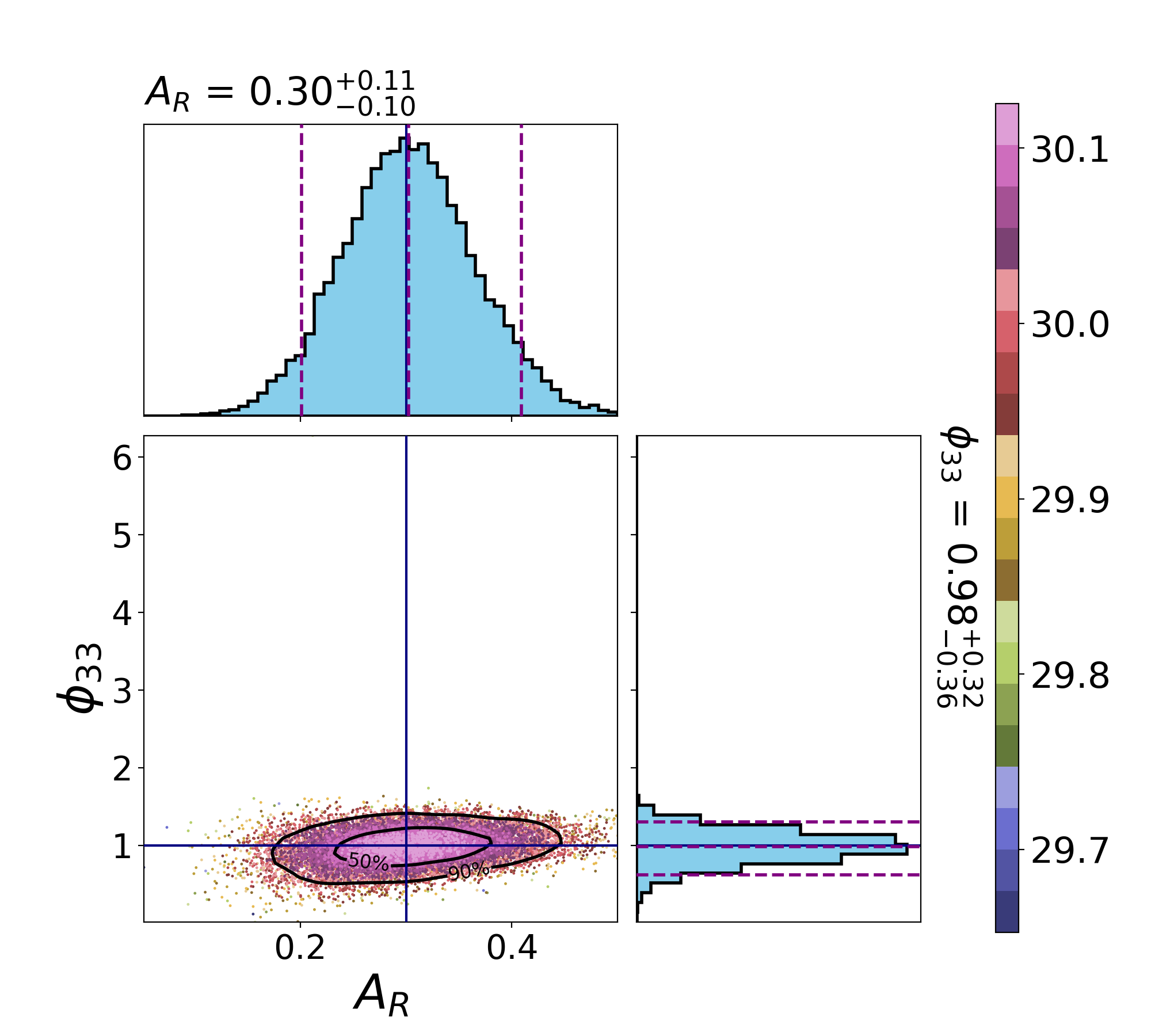} }
\caption{Posterior distributions for the amplitude ratio and phase of the subdominant mode in the $A_R = 0.3$ case. The red lines indicate the injected values. The optimal SNR of the injection is (top left) $\rho_{RD} = 15$, (top right) $\rho_{RD} = 20$, (bottom left) $\rho_{RD} = 25$, and (bottom right) $\rho_{RD} = 30$.}
\label{fig:amp_ratio_0.3}
\end{figure}

 \begin{figure}
 \subfloat{\includegraphics[width=0.25\textwidth]{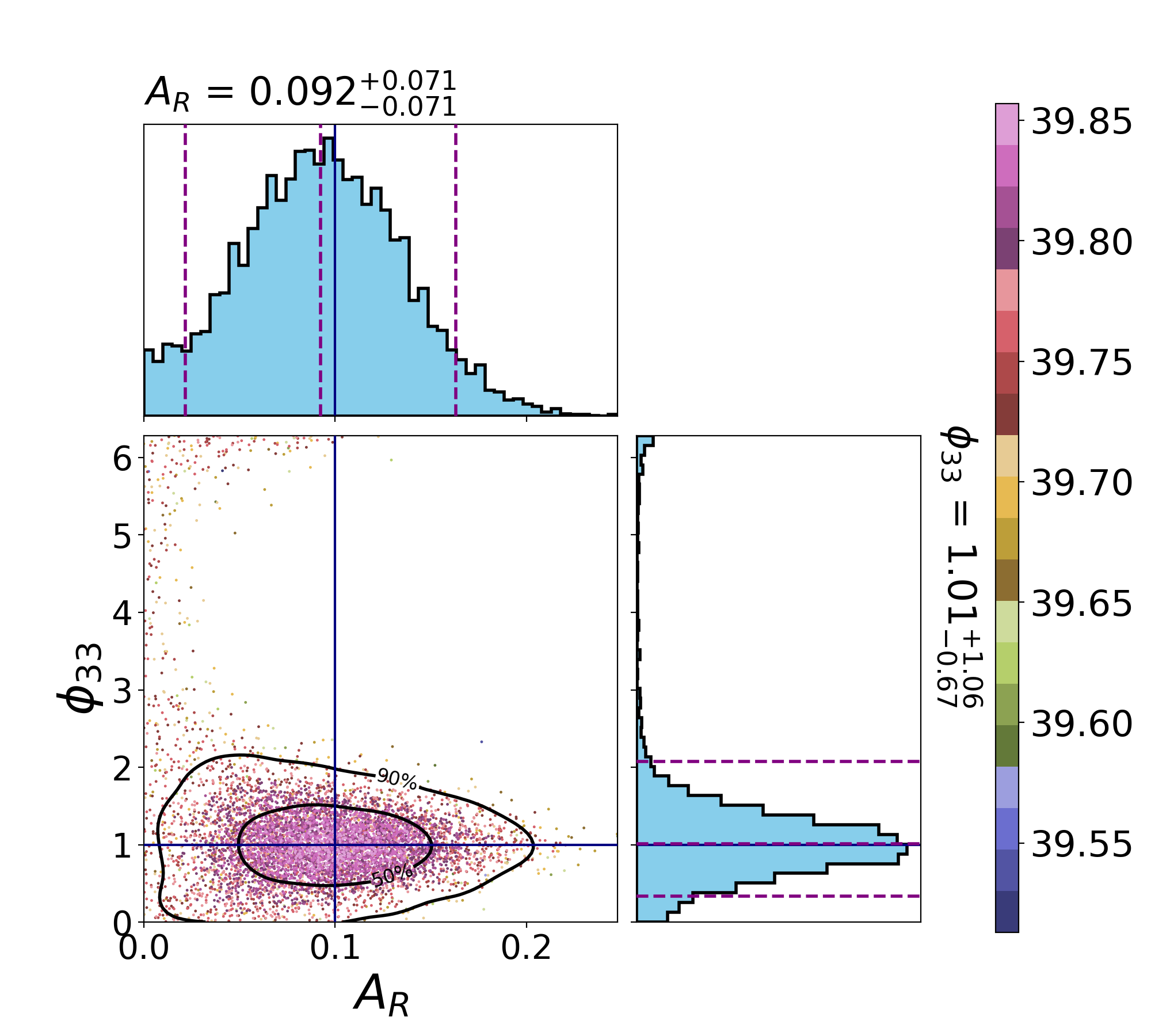} }
\subfloat{\includegraphics[width=0.25\textwidth]{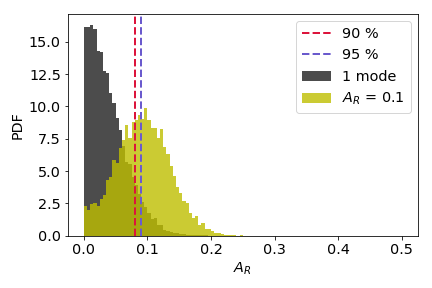} } 
\caption{Parameter estimation results for the $A_R = 0.1$ case with the SNR increased to $\rho_{RD} = 40$. (Left) Posterior distribution for the amplitude ratio and phase of the subdominant mode. (Right) Comparison of the marginalized posterior distribution of the amplitude ratio for $A_{R}=0$ (black) and $A_{R}=0.1$ (green).}
\label{fig:amp_ratio_0.3_loud}
\end{figure}
\begin{figure}
\subfloat{\includegraphics[width=0.5\textwidth]{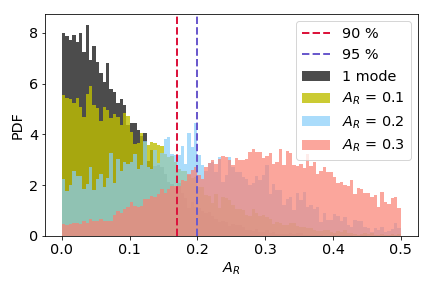} }  \\[-0.65cm]
\subfloat{\includegraphics[width=0.5\textwidth]{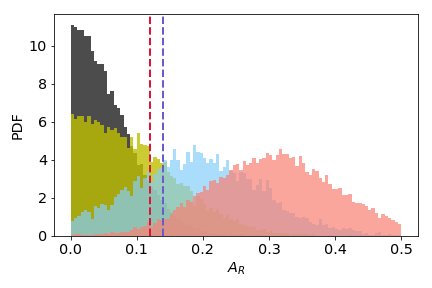} }  \\[-0.65cm]
\subfloat{\includegraphics[width=0.5\textwidth]{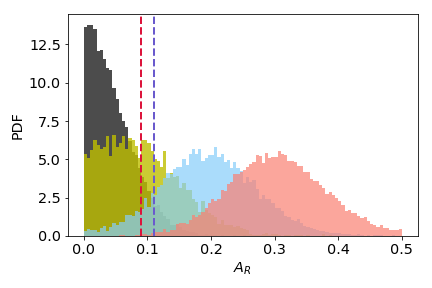} }  \\[-0.65cm]
\subfloat{\includegraphics[width=0.5\textwidth]{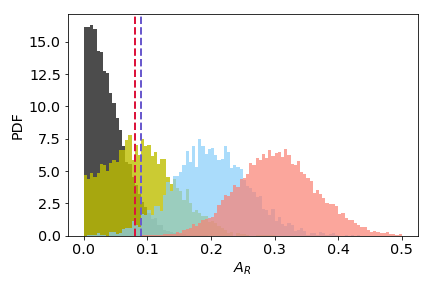} }
\caption{Marginalized posterior distributions of the amplitude ratio for all the injections in this study. From top to bottom, the ringdown injections have optimal SNR = $\{ 15, 20, 25, 30 \}$, respectively. The black histograms correspond to the null case, where the injected signal has only one mode. The green, blue and red histograms correspond to the injection with amplitude ratio of $A_R =$ 0.1, 0.2 and 0.3, respectively. We claim the detection of a second mode when the colored histograms (corresponding to a non-zero amplitude ratio) separate clearly from the black histogram.}
\label{fig:ar-posteriors}
\end{figure}
Although the parameter estimation has been performed to infer all the intrinsic parameters of the ringdown, i.e. $\{ M_{f}, a_{f}, A_{22}, A_{R}, \phi_{22}, \phi_{33} \}$, we present only the results for the inference of $ \{ A_{R}, \phi_{33} \}$ in the figures for readability. For each case the 90 \% confidence interval contains the injected values for all the parameters; an example is shown in Fig.~\ref{fig:Mf-and-af} for the mass and spin.  Figures \ref{fig:amp_ratio_0.0}, \ref{fig:amp_ratio_0.1}, \ref{fig:amp_ratio_0.2} and \ref{fig:amp_ratio_0.3} show posteriors for $A_R,\phi_{33}$ with varying $\rho_{RD}$ and $A_{R}$.
Figure \ref{fig:amp_ratio_0.0} corresponds to `the null test', where the injections contains only one mode, i.e., $A_{R} = 0$. Figures  \ref{fig:amp_ratio_0.1}, \ref{fig:amp_ratio_0.2} and \ref{fig:amp_ratio_0.3} correspond to $A_{R} =$ 0.1, 0.2 and 0.3 respectively. In all four figures, the top left panel corresponds to $\rho_{RD} = 15$, the top right panel corresponds to $\rho_{RD} = 20$, the bottom left panel corresponds to $\rho_{RD} = 25$ and the bottom right panel corresponds to $\rho_{RD} = 30$.  In each of these cases, we find that the injected value of parameters (indicated by red line in the figures) lie within 50 \% (and thus, 90 \%) credible interval. Further, the null test in Figure \ref{fig:amp_ratio_0.0} is consistent with what is expected; the marginalized posterior for $A_{R}$ rails against $A_{R} = 0$, thereby, indicating the absence of the second mode. Also, no information on the phase of $l=m=3$ mode can be inferred here. 
Among the injections we have studied, the most unlikely candidate to allow for detection of the subdominant mode is $A_{R} = 0.1$ and $\rho_{RD} = 15$ (top left panel of Figure  \ref{fig:amp_ratio_0.1}). Here we see that we cannot exclude $A_{R}=0$ from the posterior distribution, thereby not allowing us to assert the presence of the subdominant mode. Nevertheless, we notice that the posterior for $A_{R}$ has more support for higher values of $A_{R}$ compared to the null test case. As we increase the $\rho_{RD}$, the posterior distribution for $A_{R}$ shifts towards $A_{R} = 0.1$.  Since the population studies of BBH favours nearly equal mass BBH systems \cite{Abbott:2016nhf}, studying the RDs for smaller  $A_{R} $ is crucial. We therefore perform an injection with $\rho_{RD} = 40$ for the $A_{R} = 0.1$ and show that we can indeed infer the presence of the subdominant mode for a higher $\rho_{RD}$ system even with a $A_{R} = 0.1$. The PE result for this case  is presented in Figure \ref{fig:amp_ratio_0.3_loud}. 
However, it is striking that the posterior distribution for the phase of $l=m=3$ mode, $\phi_{33}$, peaks around the correct injection value even for  $A_{R} = 0.1$ and $\rho_{RD} = 15$ case. This provides strong hints favouring the presence of the subdominant mode even for  $A_{R} = 0.1$  with smaller values of $\rho_{RD}$.  We see that the phase of the sub-dominant mode is recovered better than one might expect from the distribution of $A_{R}$.

A natural criteria for a confident inference of the presence of the subdominant mode is that the 90 \% credible interval for $A_R$ should not contain any values near $A_{R} = 0$.  The credible intervals are tabulated in Table \ref{tab:result}.  With this criterion, we see that for $A_{R} = 0.1$ we cannot claim the detection of the subdominant mode at least up to $\rho_{RD} = 30$. However for $A_{R}= 0.2$, we can infer the presence of the second mode for a signal with $\rho_{RD} \geqslant 20$. When $A_{R}$ is increased to $A_{R}=0.3$, we can infer the presence of the second mode for all the injections used in this study, including  $\rho_{RD}=15$.  
\begin{table}[!htb]
 \begin{ruledtabular}
    \centering
        \begin{tabular}{ c| c| c | c }
\toprule
$\rho_{RD}$& $A_{R}$=0.1 & $A_{R}$=0.2 & $A_{R}$=0.3  \\ \hline
 15 & $6.6 \times 10^{-8} - 0.24$ & $1.4 \times 10^{-5} - 0.35$ & $ \bf{0.122 - 0.49}$  \\ \hline
 20  & $1.4 \times 10^{-5} - 0.19$ & $\bf{0.04 - 0.34}$ & $\bf{0.16 - 0.46}$\\ \hline
 25  & $5 \times 10^{-5} - 0.17$ & $\bf{0.08 - 0.32}$ & $\bf{1.18 - 0.43}$ \\ \hline
 30  & $\bf{1.2 \times 10^{-4} - 0.16}$ & $\bf{0.1 - 0.3}$ & $\bf{0.2 -0.4}$ \\ 
 \bottomrule
\end{tabular}
        \caption{90 \% highest posterior density (HPD) credible interval on the marginalized PDF of $A_{R}$. In $\bf{bold}$ are the cases where we are able to infer the presence of the subdominant mode. Posterior distributions are shown in Figures \ref{fig:amp_ratio_0.0}, \ref{fig:amp_ratio_0.1}, \ref{fig:amp_ratio_0.2} and \ref{fig:amp_ratio_0.3}}
\label{tab:result}
\end{ruledtabular}
\end{table}

A different method is to compare the marginalized posteriors distributions for $A_{R}$  for different SNR with the null case (this will allow us to get an intuition for the false alarm and false dismissal probabilities for the inferred presence of subdominant mode).  Figure \ref{fig:ar-posteriors} shows the marginalized posterior distributions for $A_R$ where the panels are arranged top to bottom for $\rho_{RD} = 15, 20, 25, 30$ respectively.  To define the false dismissal probability $\beta$, we need to choose a threshold on $A_R$ based on the null case.  The thresholds $A_R^{90\%}$ and $A_R^{95\%}$ corresponding respectively to $90\%$ and $95\%$ false alarm rates are shown as vertical lines in Figure \ref{fig:ar-posteriors}.  For any of these thresholds, say $A_R^{90\%}$, the false dismissal probability is
\begin{equation}
    \beta^{90\%} = \int_0^{A_R^{90\%}} p(A_R|\widehat{A}_R)dA_R
\end{equation}
where $\widehat{A}_R$ is the true injected value of $A_R$.  
If the posterior distribution for $A_{R}$ clearly separates from the posterior distribution for the case with $A_{R}=0$, the presence of the second mode can be inferred confidently. We note that the posterior distribution corresponding to $A_{R} = 0.3$ (the pink histogram) always separates from $A_{R} = 0 $ (the black histogram), even for $\rho_{RD}=15$, whereas that which corresponds to  $A_{R}=0.2$  (the blue histogram) separates out after $\rho_{RD} = 20$.   Table \ref{tab:beta} lists the values of $\beta$ for the various cases.   These results are consistent with the credible intervals listed in Table \ref{tab:result}.  Finally, in the left panel of Figure \ref{fig:amp_ratio_0.3_loud}, we do see that for $\rho_{RD} = 40$, we can infer the presence of subdominant mode in case of $A_{R} = 0.1$ confidently.

\begin{table}[]
 \begin{ruledtabular}
        \begin{tabular}{ c| c | c | c }
\toprule
$\rho_{RD}$& $A_{R}$=0.1 & $A_{R}$=0.2 & $A_{R}$=0.3  \\ \hline
 15 & $0.75$ & $0.46$ & $0.17$  \\ \hline
 20  & $0.67$ & $0.23$ & $0.02$\\ \hline
 25  & $0.53$ & $0.07$ & $1.6 \times 10^{-3}$ \\ \hline
 30  & $0.45$ & $0.02$ & $10^{-4}$ \\ 
 \bottomrule
\end{tabular}
        \caption{False dismissal probability $\beta^{90\%}$ for detection of a non-zero value of $A_{R}$ for different values of the SNR and injected amplitude $A_R$.}
\label{tab:beta}
\end{ruledtabular}
\end{table}

\section{Discussions and Implications}
\label{sec:conclusion}

The realization of black hole spectroscopy, i.e. using the ringdown spectrum of a Kerr black hole to test the black hole no-hair theorem, requires the detection of multiple ringdown modes.  In this paper we have applied Bayesian inference techniques to the problem of detecting a sub-dominant ringdown mode for different values of the total SNR $\rho_{RD}$ and the amplitude ratio $A_R$. In our study we have also taken the $\ell=m=2, n=0$ and $\ell=m=3,n=0$ modes to be the dominant and sub-dominant modes respectively.  It would be straightforward to extend include other choices for the sub-dominant modes, including higher overtones.  For the two LIGO detectors operating at design sensitivity, we find that for a remnant black hole with mass and spin similar to GW150914, sub-dominant modes with $A_R=0.1$ are potentially detectable for $\rho_{RD} = 30$, and amplitude ratios of $0.3$ would be detectable at $\rho_{RD} = 15$.

For a BBH ringdown signal, the excitation amplitude of the different modes depends on the perturbation conditions set up by the inspiral-merger phase. This, in turn, is dictated by the asymmetry of the initial binary system i.e., mass ratio and the spin of the progenitor. Generally speaking, more asymmetric systems will have higher modes excited, but are also less likely to be detected.  This is already clear from the current list of binary black hole mergers which have been detected \cite{catalogue1,catalogue2,catalogue3}. The general question of how likely we are to detect a sufficiently asymmetric system that can be detected with networks of future gravitational wave detectors, and the important question of determining the frequencies and damping times rather than just detecting them, will be addressed in a companion paper \cite{LIGOScientific:2018jsj}.

The premise of this study is to only `detect' the presence of the second mode in RD under the assumption that the underlying theory of gravity is GR. Deriving the QNM frequencies from the final mass and spin of the BH instead of leaving them as free parameters to be inferred during the PE, reduces the parameter space by two parameters, thereby, allowing the detection of the second mode at a smaller SNR. Note, however, that if there is a deviation of the signal from what is predicted by GR, it is expected to be reflected as features of posteriors inferred for the final mass and final spin of the final BH; for example, one might expect to observe features like multimodal posterior for final mass and spin if the frequency for the second mode is significantly different from the GR predictions. Also, given the SNR  comparing the variance of the inferred posteriors with what would be expected for a GR case could provide hints towards possible violations of GR.  Although the setup we have used already sheds some light on our assumption that the underlying theory is GR, a more robust test of GR would require measurement of the QNM frequencies and damping times directly from the data. Validating theorems like no hair theorem require the measurement of the QNM spectrum.  A rough estimate of SNR required to measure the frequencies can be calculated using a Fisher matrix framework combined with the Rayleigh criterion as presented in \cite{bertiparam}. From this rough calculation, we expect that for an amplitude ratio $A_{R} = \{ 0.1, 0.2, 0.3 \}$, we will require $\rho_{RD} \sim 25, 13, 9$ respectively.  Measurement of subdominant mode frequencies and damping time in a fully Bayesian framework is currently one of our ongoing studies.

\section{Acknowledgements}
\label{sec:acknowledgements}

We would like to thank Alexander H. Nitz and Steven Reyes for useful discussions. SB and DAB thank National Science Foundation Award No.~PHY-1707954 for support. SB acknowledges financial support provided under the European Union’s H2020 ERC, Starting Grant agreement no.
DarkGRA–757480 and networking support by the COST Action CA16104 and
from the Amaldi Research Center funded by the MIUR program Dipartimento di Eccellenza (CUP: B81I18001170001). MC acknowledges support from National Science Foundation Award No.~PHY-1607449, the Simons Foundation, and the Canadian Institute For Advanced Research (CIFAR). Computations were supported in part through computational resources provided by Syracuse University, supported by National Science Foundation Grant No.~ACI-1541396, and by the Atlas computer cluster at the Albert Einstein Institute (Hannover).





\bibliography{biblo}
\end{document}